\documentclass[aps,prb,twocolumn,superscriptaddress,showpacs,floatfix,amsmath,amssymb,10pt]{revtex4-1}
\usepackage{graphicx}
\usepackage{float}
\usepackage{caption}
\usepackage{amsfonts}
\usepackage{physics}

\begin{document}
 
\captionsetup[figure]{calcwidth=0.8\linewidth}

\title{InfoCGAN Classification of 2-Dimensional Square Ising Configurations}

\author{Nicholas Walker}
\affiliation{Department of Physics \& Astronomy, Louisiana State University, Baton Rouge, Louisiana 70803, USA}
\author{Ka-Ming Tam}
\affiliation{Department of Physics \& Astronomy, Louisiana State University, Baton Rouge, Louisiana 70803, USA}
\affiliation{Center for Computation \& Technology, Louisiana State University, Baton Rouge, Louisiana 70803, USA}

\begin{abstract}

An InfoCGAN neural network is trained on 2-dimensional square Ising configurations conditioned on the external applied magnetic field and the temperature. The network is composed of two main sub-networks. The generator network learns to generate convincing Ising configurations and the discriminator network learns to discriminate between ``real'' and ``fake'' configurations with an additional categorical assignment prediction provided by an auxiliary network. Some of the predicted categorical assignments show agreement with the expected physical phases in the Ising model, the ferromagnetic spin-up and spin down phases as well as the high temperature weak external field phase. Additionally, configurations associated with the crossover phenomena are predicted by the model. The classification probabilities allow for a robust method of estimating the critical temperature in the vanishing field case, showing exceptional agreement with the known physics. This work indicates that a representation learning approach using an adversarial neural network can be used to identify categories that strongly resemble physical phases with no \textit{a priori} information beyond raw physical configurations and the physical conditions they are subject to.

\end{abstract}

\maketitle

\section{Introduction}

Recent advances in the field of machine learning have provided many opportunities for researching applications in the sciences. Fundamentally, the machine learning approach is motivated by the use of pattern recognition and statistical inference in a manner capable of isolating statistically significant features to produce meaningful predictions. This is in contrast to more traditional methods that require explicit instruction sets to elicit meaningful predictions from input data. In many cases, the appropriate instructions are not necessarily known, which renders a machine learning approach attractive for addressing such systems. Generally speaking, obtaining a prediction using a feed-forward artificial neural network involves tuning the weights and edges in a computational graph to minimize error in the network given an input and an output. Interestingly, it has been shown that there is an exact mapping between the variational renormalization group (RG) and the RBM, perhaps implying that machine learning approaches may employ a generalized RG-like scheme to extract relevant features from input data \cite{ml_rg}.

Identifying statistically significant features in physical systems to determine criteria for phase transitions is a well-known approach. One example is the widely-adopted Lindemann parameter, which is often used to characterize the melting of crystal structures by measuring the deviations of atoms from equilibrium positions \cite{lindemann}. For sufficiently complex systems, the identification of such a criterion can be inaccessible through traditional means, which opens an opportunity for machine learning approaches to accomplish such a task. This concept of hidden orderings has been proposed for some systems \cite{hel_ord,cup_hid_ord,hf_hid_ord}. Although inference methods, such as the maximum likelihood method and the maximum entropy method \cite{max_like,max_ent} have been routinely applied on certain physical problems, applications which utilize other machine learning methods have not attracted much attention until recently. A lot of work has been done in recent times for detecting phase transitions and performing structure classifications in various physical systems using machine learning \cite{ml_pom,ml_pt,struc_clss,ml_pt_lat,frust_spin_1,frust_spin_2,melting}. Additionally, some exciting work has been done to investigate the capacity for machine learning methods to accelerate numerical simulations of physical systems \cite{acc_mc}. For the Ising system specifically, there have been many such efforts. Boltzmann machines (BMs), deep belief networks (DBNs), and deep restricted Boltzmann machines (RBMs) have proven to be effective at generating Ising configurations with accurate thermodynamic properties \cite{ising_boltzmann,ising_boltzmann_2}. Variational autoencoders (VAEs) have also been used to successfully extract the Ising order parameter from the predicted latent representations in the vanishing field case \cite{ising_vae,ising_vae_2,ising_vae_3}. Additionally, it has been shown that supervised classifier neural networks are indeed capable of providing an accurate estimation of the Ising critical point in the vanishing field case \cite{ising_class_small,ising_class_order}. Accurate classification can even be obtained with a rather small network and the order parameter can be even be extracted from the decision function of the classification network, albeit through the use of supervised learning with \textit{a priori} categorization labels.

Focus in this work will be placed on the 2-dimensional square Ising model, an oft-employed model for describing magnetic phenomena in the field of statistical mechanics. The model is notable for describing one of the simplest systems in statistical physics that demonstrates a phase transition. In this case, a second-order magnetic transition between the ferromagnetic and paramagnetic phases at the Curie temperature (also referred to as the critical point in the context of a second-order phase transition). The model has seen extensive use in investigating magnetic phenomena in condensed matter physics \cite{spin_glass_1,spin_glass_2,ising_app_1,ising_app_2,ising_app_3,ising_app_4,ising_app_5}. Additionally, the model can be equivalently expressed in the form of the lattice gas model. This is a simple model of density fluctuation and liquid-gas transformations used primarily in chemistry, albeit often with modifications \cite{ising_chem_1,ising_chem_2}. Furthermore, modified versions of lattice gas models have been applied to studying binding behavior in biology \cite{ising_bio_1,ising_bio_2,ising_bio_3}.

The purpose of this work is to explore the capability of machine learning to discriminate between different classes of microstates (the possible microscopic configurations of the thermodynamic system) exhibited by the 2-dimensional square Ising model in an unsupervised manner. The properties of these classes will then be compared to the expected properties of the ferromagnetic and paramagnetic phases, as well as the crossover regions. The ability to use unsupervised machine learning to detect phases of matter carries broad implications for possible applications in the study of physical systems with unknown phase diagrams. Furthermore, the detection of crossover regions opens a new avenue for the study of quantum critical points, where numerical data must be obtained for low, but finite temperatures that exhibit crossover instead of criticality. Examples include data from large scale numerical quantum Monte Carlo simulations for heavy fermion materials and high temperature superconducting cuprates for which quantum critical points are believed to play crucial roles for their interesting properties \cite{qcp,singular_fl,2D_DCA_QCP}. While prior research has utilized either principal component analysis (PCA) or VAEs to perform unsupervised representation learning of Ising configurations \cite{ising_vae,ising_vae_2,ising_vae_3}, this work is taking an alternative approach through the use of an information maximizing conditional generative adversarial network (InfoCGAN) \cite{infocgan}. The prior research has been focused on the extraction of particular physical properties of the Ising configurations such as the magnetization. The magnetization quantifies the order exhibited by the system across the phase transition and serves as a discrimination criterion between physical phases. Thus, such an approach is consistent with traditional approaches in statistical mechanics. By contrast, the InfoCGAN approach pursued in this research uses a direct classification scheme aided by conditioning the network on external physical conditions to predict a categorical latent space. This is similar to other research that has been done to classify Ising configurations in order to predict the critical point \cite{ising_class_small,ising_class_order}, but these approaches utilize supervised learning that requires \textit{a priori} knowledge of the physical phases the configurations belong. By contrast, the research in this work utilizes unsupervised learning that does not require any prior knowledge of physical phases. In effect, the InfoCGAN approach provides a direct unsupervised phase classification scheme that may be useful for analyzing systems in which an order parameter that cleanly translates to a classification criterion is not obvious. Furthermore, this approach grants additional capability for conditioning the neural network on additional known parameters that describe not just the Ising configurations, but the physical conditions the configurations are subject to, which should improve modeling of the distribution of the training data \cite{cgan}. Additionally, the adversarial loss function utilized by the InfoCGAN model as an optimization objective serves as a feature-wise loss function that has been shown to provide greater visual fidelity in outputs for computer vision applications when compared to those obtained with the element-wise reconstruction loss functions used in VAEs \cite{elem_feat}. Given that 2-dimensional square Ising configurations can be readily interpreted as images, it is reasonable to expect that an adversarial approach will perform well in the context of unsupervised learning of the Ising model by comparison to prior research that has relied on VAEs.

\section{The Ising Model}

The Ising model is mathematically expressed in the form of a 2-dimensional array of spins $\sigma_i$ indexed by $i$ that represent a discrete arrangement (lattice) of dipole moments of atomic spins (the magnetic orientation of the lattice sites) \cite{p_cond_mat}. The spins are restricted to spin-up or spin-down alignments represented as $\pm 1$ such that $\sigma_i \in \Bqty{-1,+1}$. The spins are allowed to interact with their nearest neighbors according to the pair-wise interaction strength $J_{ij}$ for neighbors $\sigma_i$ and $\sigma_j$. Additionally, each spin $\sigma_i$ will interact with an external magnetic field $H_i$ where the magnetic moment $\mu$ has been absorbed into the magnetic field $H_i$ at that lattice site. The magnetic moment is an expression of the magnetic strength and orientation of a material and serves to control the interaction strength of the lattice sites with the external field. The full Hamiltonian (an expression of the total energy) describing this system is thus expressed as

\begin{equation}
\mathcal{H} = -\sum_{\expval{i,j}}J_{ij}\sigma_i\sigma_j-\sum_i H_i\sigma_i.
\end{equation}

Where $\expval{i,j}$ indicates adjacent index pairs $i$ and $j$. For $J_{ij} > 0$, the interaction between spins $i$ and $j$ is ferromagnetic, for $J_{ij} < 0$, the interaction between spins $i$ and $j$ is antiferromagnetic, and for $J_{ij} = 0$, the spins $i$ and $j$ are necessarily non-interacting. Furthermore, for $H_i > 0$, the spin at lattice site $i$ tends to prefer a spin-up alignment, with the strength of the preference determined by the strength of the field. By contrast, for $H_i < 0$, the spin at lattice site $i$ tends to prefer a spin-down alignment. For $H_i = 0$, there is no external magnetic field influence on lattice site $i$. Typically, the Ising model is solved for the case $J_{ij} = J$ and $H_i = H$. This is the expression used in this work as follows

\begin{equation}
\mathcal{H} = -J\sum_{\expval{i,j}}\sigma_i\sigma_j-H\sum_i\sigma_i.
\end{equation}

The Ising model can also be re-expressed in the form of a lattice gas. The resulting modified Hamiltonian is represented in the following manner

\begin{equation}
\mathcal{H} = -4J\sum_{\expval{i,j}}n_i n_j - \mu\sum_i n_i.
\end{equation}

Where the external field strength $H$ is reinterpreted as the chemical potential $\mu$, $J$ retains its role as the interaction strength, and $n_i \in \Bqty{0, 1}$ represents the lattice site occupancy in which a lattice site may contain an atom (1) or not (0). The original Ising Hamiltonian can be recovered using the relation $\sigma_i = 2n_i-1$ up to a constant. In this model, the tendency for adjacent spins to align with one another can now be re-interpreted as the tendency for atoms on a lattice to be attracted to one another. Additionally, the chemical potential provides the energy cost associated with adding an atom to the system from a reservoir. This form of the model has many applications in chemistry and biology, as mentioned in the introduction.

The Ising spin configurations are Boltzmann distributed according to the energy and temperature of the system, which can be expressed as

\begin{equation}
p_n = \frac{e^{-\beta E_n}}{Z};\quad Z = \sum_{n} e^{-\beta E_n};\quad \beta = \frac{1}{T}.
\end{equation}

Where $p_n$ is the probability of a state existing in state $n$ at temperature $T$, $E_n$ is the energy of state $n$, $Z$ is the normalization constant referred to as the partition function, and $\beta$ is the inverse temperature. The Boltzmann constant $k_B$ was absorbed into the temperature $T$.

For the Ising model, the vanishing field case $H=0$ is of particular interest. Under these conditions, the model exhibits a second-order phase transition characterized by a critical temperature $T_{C}$ for dimension $d\ge 2$. The 2-dimensional case was originally solved analytically by Onsager in 1944 \cite{onsager}. At low temperatures below the critical point, the system is dominated by nearest-neighbor interactions, which for a ferromagnetic system means that adjacent spins tend to align with one another. As the temperature is increased, thermal fluctuations will eventually overpower the interactions such that the magnetic ordering is destroyed and the orientations of adjacent spins can be taken to be uncorrelated. This is referred to as a paramagnet which will still tend to align with an external field. However, for sufficiently high temperatures, a sufficiently strong external field will be required to overcome the effects of thermal fluctuations. The ferromagnetic and paramagnetic phases are discriminated by an order parameter, which takes the form of the magnetization for the ferromagnetic Ising model. Oftentimes, the average magnetization is used to fulfill this role and for a system composed of $N$ lattice sites, it is expressed as

\begin{equation}
m = \frac{1}{N}\sum_{i=1}^N \sigma_i.
\end{equation}

In the vanishing field case for a system in the thermodynamic limit ($N \rightarrow \infty$, $V \rightarrow \infty$), the average magnetization vanishes at the critical temperature with power law decay behavior $m(T) \sim (T_C-T)^\beta$ with critical exponent $\beta = \frac{1}{8}$ for two dimensions. The magnetic susceptibility $\chi$ and the specific heat capacity $C$ additionally diverge at the critical temperature. They are expressed as follows

\begin{equation}
\chi = \frac{\expval{m^2}-\expval{m}^2}{T};\quad C = \frac{\expval{E^2}-\expval{E}^2}{T^2}.
\end{equation}

The system can be said to be paramagnetic for a vanishing average magnetization and ferromagnetic for a finite average magnetization (with the system categorized as ferromagnetic spin-up or spin-down according to the sign of the magnetization). However, in the presence of a non-vanishing external magnetic field, an alignment preference is introduced according to the sign of the field, which destroys the $Z2$ symmetry. As a result, a first-order phase transition is observed at low temperatures ($T < T_C$) by varying the strength of the magnetic field from negative to positive values (or vice versa), which produces a discontinuous change in the average magnetization across the vanishing field line. Furthermore, in the presence of an external field, there is no longer a second-order phase transition, as the average magnetization no longer vanishes at a critical temperature. Rather, there is a region in which the system tends towards a disordered state from an ordered state. This region is referred to as the crossover region, which is not a thermodynamic phase. Generically, a crossover refers to when a system undergoes a change in phase without encountering a canonical phase transition characterized by a critical point as there are no discontinuities in derivatives of the free energy (as determined by Ehrenfest classification) or symmetry-breaking mechanisms (as determined by Landau classification). A well known example is the BEC-BCS crossover in an ultracold Fermi gas in which tuning the interaction strength (the s-wave scattering length) causes the system to crossover from a Bose-Einstein-condensate state to a Bardeen-Cooper-Schrieffer state \cite{bec_bcs}. Additionally, the Kondo Effect is important in certain metallic compounds with dilute concentrations of magnetic impurities that cross over from a weakly-coupled Fermi liquid phase to a local Fermi liquid phase upon reducing the temperature below some threshold \cite{kondo}. Furthermore, examples of strong crossover phenomena have also been recently discovered in classical models of statistical mechanics such as the Blume-Capel model and the random-field Ising model \cite{crossover_bcm,crossover_rfim}. Additionally, for a finite-sized system, there is no true critical point since the system is not evaluated in the thermodynamic limit.

\begin{figure}[H]
\centering
\includegraphics[width=0.6\linewidth]{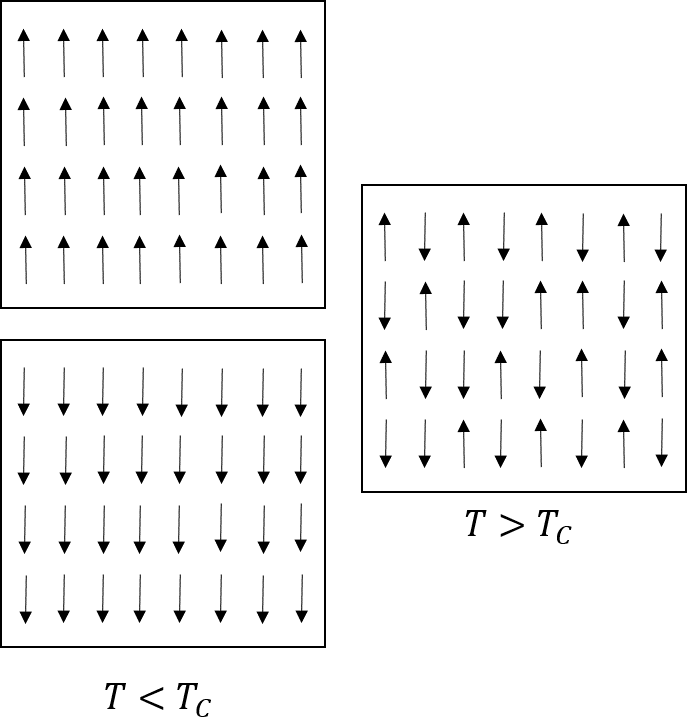}
\caption{\footnotesize This diagram depicts the spin configurations in $2$ dimensions.}
\label{fig:spins}
\end{figure}

\begin{figure}[H]
\centering
\includegraphics[width=0.6\linewidth]{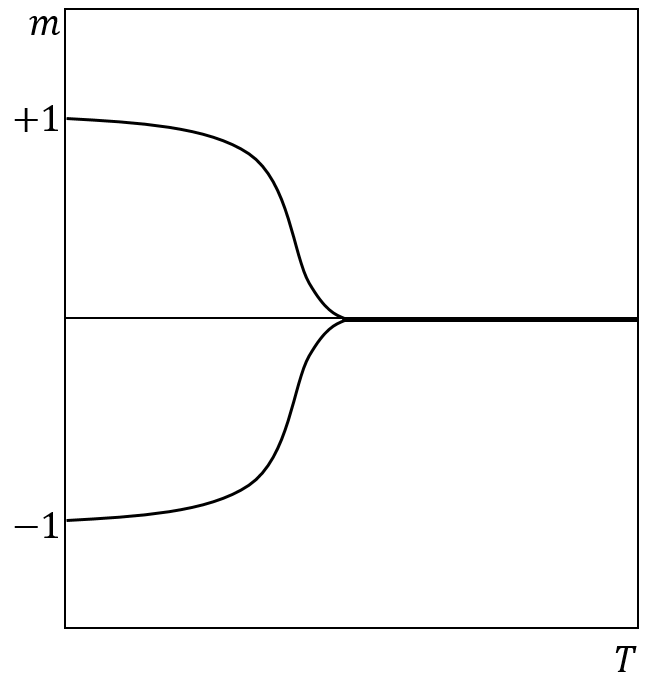}
\caption{\footnotesize This diagram depicts a sketch of the temperature dependence of the magnetization in the vanishing field case.}
\label{fig:mag_v}
\end{figure}

\begin{figure}[H]
\centering
\includegraphics[width=0.6\linewidth]{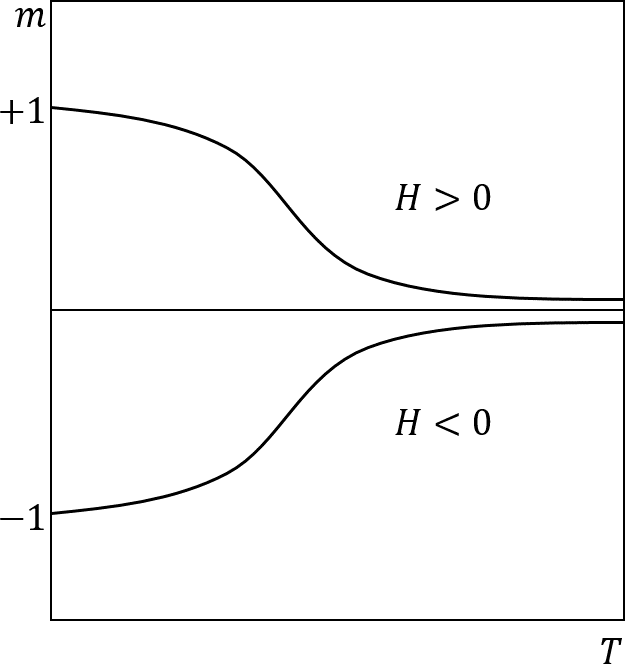}
\caption{\footnotesize This diagram depicts a sketch of the temperature dependence of the magnetization in the non-vanishing field case.}
\label{fig:mag_nv}
\end{figure}

Examples of the types of spin configurations that can be found in the vanishing field case are depicted in Fig.\ref{fig:spins}. Below the critical temperature, nearest neighbor interactions dominate the system, causing the spins to align with one another in either the spin-up or spin-down configuration. The behavior of the average magnetization as a function of temperature is additionally depicted for the vanishing field case and the non-vanishing field case in Fig.\ref{fig:mag_v} and Fig.\ref{fig:mag_nv} respectively. For the vanishing field case, the average magnetization can be seen to vanish in the vicinity of the critical temperature. By contrast, the average magnetization tends to smoothly decay to a small value in the non-vanishing field case, with sign determined by the external magnetic field.

\section{Generative Adversarial Networks and the InfoCGAN Model}

The InfoGAN class of artificial neural networks belong to the larger group of generative adversarial networks (GAN) and are thus guided by the same principles \cite{infogan,gan}. A standard GAN is composed of two primary components, a generator network and a discriminator network, implemented as computational graphs with many adjustable parameters \cite{gan}. In this framework, the networks are interpreted to be contesting with one other in an adversarial game (albeit not always a zero-sum game) to solve a non-convex optimization of loss in the network outputs. The non-convex optimization problem arises in the high-dimensional space created by the large number of parameters within the neural networks, which tends to produce potentially many local minima, saddle points, flat regions, and widely varying curvature on the loss surface. As such, non-convex optimization can be rather difficult. From the perspective of an adversarial game, the discriminator learns to distinguish between ``real'' samples drawn from the distribution of the training data and ``fake'' samples provided by the generator. At the same time, the generator learns to ``fool'' the discriminator with ``fake'' samples. During training, the generator will learn to produce ``fake'' samples that increasingly resemble the ``real'' samples by using information provided by the training of the discriminator. More specifically, the discriminator network transforms an input sample into a measure of the probability that it belongs to the same statistical distribution as the training samples. The output of the discriminator network is a single value belonging to the interval $\hat{v} \in \qty[0,1]$ which is referred to as the validity of the sample. The magnitude of the validity score represents the probability that the input sample belongs to the same distribution as the training samples. The generator network transforms an input Gaussian noise vector into a sample that ideally belongs to the same statistical distribution as the training samples. The output of the generator network is thus of the same shape as the training samples.

The procedure for training a GAN is done by first optimizing the weights of the discriminator network to predict a batch (subset) of the training samples as ``real'' (an output validity score of one) and a batch of samples obtained from the generator network as ``fake'' (an output validity score of zero). Then, the generator network weights are optimized such that given a batch of random Gaussian latent variables, the discriminator outputs a validity score corresponding to a ``real'' sample (with the discriminator weights held constant during this step). This procedure is repeated for each batch in the training set to constitute a training epoch. In this manner, the generator and the discriminator networks ``learn'' from each other, as the discriminator network is optimized to detect differences between the ``real'' training samples and the ``fake'' generated samples while the generator network is optimized to respond to information provided by updates to the discriminator network to generate more convincing ``fake'' samples that the discriminator network identifies as ``real.'' For the implementation used in this work, the minmax GAN (or MM GAN), the adversarial contest between the discriminator and generator networks is indeed a zero-sum game in which gains and losses are balanced between the networks. As with most feedforward artificial neural networks, the tuning of the weights in the networks is done by performing backpropagation to minimize an objective loss function. This is done by calculating the gradients of the loss functions with respect to the parameters in the networks in order and updating them according to the direction of greatest descent in model loss. This is implemented with a chosen optimizer. The ideal outcome of the training process for an MM GAN is then to produce a stable Nash equilibrium between the discriminator and generator networks in which the state of the objective loss function ensures that the networks will not change their behavior regardless of what its opponent may do. For the specific case of the MM GAN, the cost function (sum of the model losses) takes the following form

\begin{align}
\min_{G}\max_{D} \mathcal{V}_{\mathrm{GAN}}(D, G) = &\mathbb{E}_{x\sim P(x)}\qty[\log D(x)]+\\\nonumber
&\mathbb{E}_{z\sim P(z)}\qty[\log\qty(1-D(G(z)))].
\end{align}

Where $x$ represents a training configuration, $z$ represents latent input to the generator network, $P(x)$ represents the distribution of the training samples, $P(z)$ represents the distribution of the latent input to the generator network, $D$ represents the discriminator network, and $G$ represents the generator network. This can be reinterpreted as objective loss functions for the respective networks in the following manner

\begin{align}
\mathcal{L}_D &= -\mathbb{E}_{x\sim P(x)}\qty[\log D(x)]-\mathbb{E}_{\hat{x}\sim P(\hat{x})}\qty[\log\qty(1-D(\hat{x}))] \\
\mathcal{L}_G &= \mathbb{E}_{\hat{x}\sim P(\hat{x})}\qty[\log\qty(1-D(\hat{x}))].
\end{align}

Where now $\hat{x}$ has been adopted to represent a generated sample drawn from the distribution $P(\hat{x})$ provided by the generator network. Training a GAN is notoriously difficult, as the model is sensitive to many major problems. These including simple non-convergence to a Nash equilibrium, mode collapse in which the generator fails to produce a variety of samples, diminished gradients in which vanishing gradients from an effective discriminator fails to provide enough information to the generator, imbalance between the generator and discriminator resulting in overfitting, and sensitivity to training hyperparameters, to name a few.

With the motivations behind GANs in mind, it is clear to see the value that such a framework can provide to researching physical systems. Assuming a GAN is sufficiently trained on configurations drawn from a well-simulated physical system, the distribution of the microstates composing the simulation data can be ``learned,'' which can then be used to predict physical properties of interest. This is not unlike calculating physical properties from the partition function describing a system in statistical physics. However, while the standard GAN framework provides the necessary tools for generating convincing configurations as well as discriminating between ``real'' and ``fake'' ones, this is not necessarily of immediate interest to physical research. Indeed, there is an interpretability problem with GANs. While the important information required to draw a sample from the distribution of the training configurations is ostensibly contained in the latent input to the generator network, there is not always an obvious way to extract desired properties from the description provided by the latent input. That is to say that a standard GAN does not provide disentangled representations in the latent space from which clear features of the data can be interpreted, such as facial expression from computer vision applications. The InfoGAN model provides an encouraging attempt to address this interpretability problem by learning disentangled representations of configurations by maximizing mutual information between small subsets of the latent inputs to the generator network and the output of an auxiliary network to the discriminator network \cite{infogan}. 

This is done by including an additional interpretable input $c$ (often referred to as the control codes) alongside the random latent input $z$ to the generator network in order to produce a configuration $\hat{x}$. Through the auxiliary network, inputs to the discriminator network provided by the generator network $\hat{x}$ are mapped to a prediction of the control codes $\hat{c}$ using the same extracted features that the discriminator uses to determine the validity $\hat{v}$ of the input. The auxiliary network is then optimized to minimize the loss between $c$ and $\hat{c}$. Thus, interpetable control codes can be predicted for configurations from the training set through the auxiliary network. This can be incorporated into the cost function when training the discriminator network through an additional mutual information term $I(c; G(c, z))$ where $c$ distributed as $P(c)$ with entropy $H(c)$, $G(c, z)$ is the generator network, and $Q(c|x)$ is the auxiliary network. The mutual information term is estimated using an entropic approach. The InfoGAN cost function is expressed as

\begin{align}
\min_{G}\max_{D} \mathcal{V}_{\mathrm{InfoGAN}}(D, G) &= \mathcal{V}_{\mathrm{GAN}}(D, G) - \lambda I(c; G(c, z));\\
I(c; G(c, z)) &= H(c)-H(c|G(c, z)) \\\nonumber
&\ge \mathbb{E}_{c\sim P(c), x\sim G(c, z)}\qty[\log Q(c|x)]+\\
\nonumber
&\quad H(c).
\end{align}

Where $\lambda$ is a hyperparameter used to weigh the importance of optimizing the mutual information against the generative ability of the model to produce ``real'' samples and the discriminative ability of the model to distinguish ``real'' and ``fake'' samples. The control codes $c$ are free to take on many forms, but are most frequently taken to be uniform, Gaussian, or categorical variables. In this work, categorical variables are employed to allow the network to directly classify different kinds of Ising configurations in an unsupervised manner. The use of conditional variables is not restricted to these control variables, however.

\begin{figure}[H]
\centering
\includegraphics[width=0.8\linewidth]{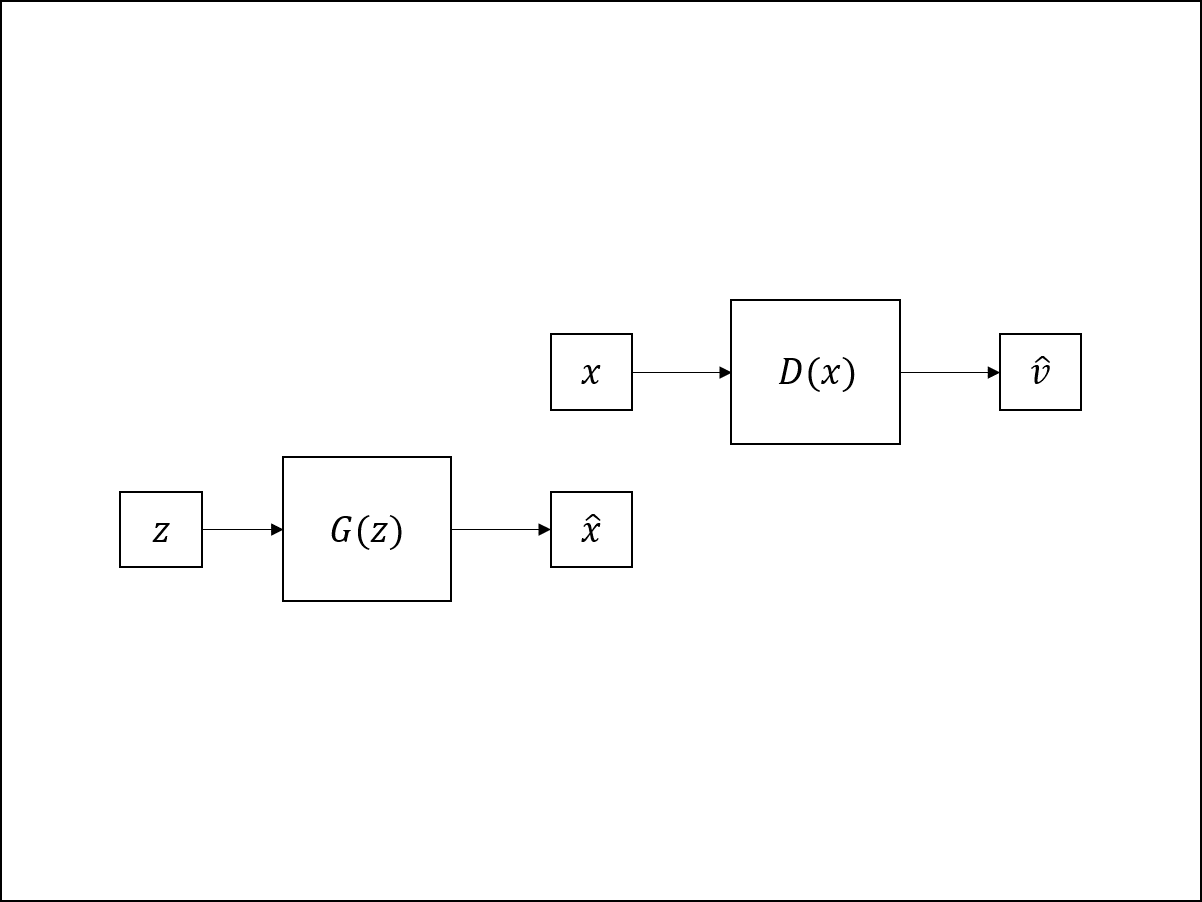}
\caption{\footnotesize This diagram depicts the structure of a GAN model.}
\label{fig:gan}
\end{figure}

\begin{figure}[H]
\centering
\includegraphics[width=0.8\linewidth]{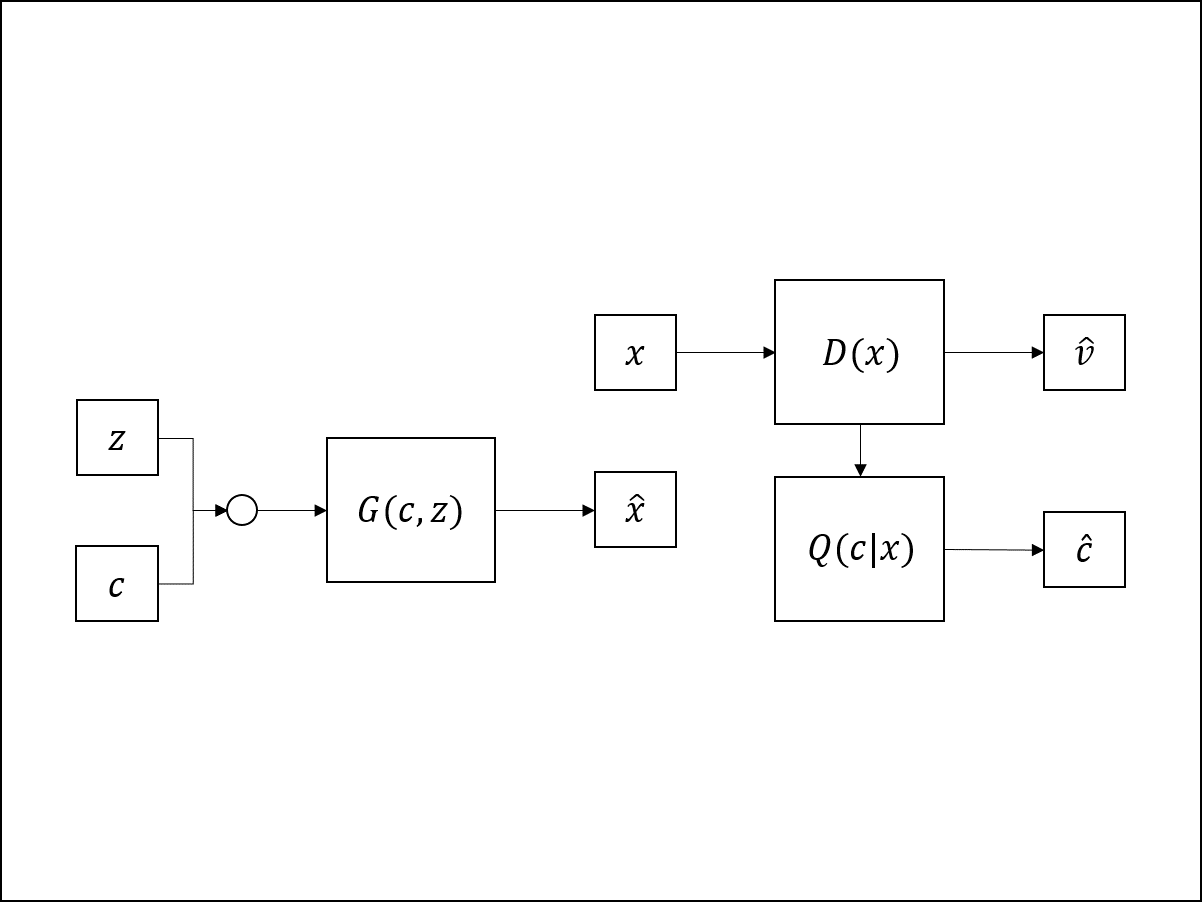}
\caption{\footnotesize This diagram depicts the the structure of an InfoGAN model.}
\label{fig:infogan}
\end{figure}

The structures described for the GAN and InfoGAN models are respectively depicted in Fig.\ref{fig:gan} and Fig.\ref{fig:infogan}. For the GAN, the generator model $G(z)$ takes a noise vector $z$, which is usually normally distributed, as an input to generate a ``fake'' sample $\hat{x}$. The discriminator model $D(x)$ then takes either a ``real'' sample $x$ or a ``fake'' sample $\hat{x}$ as an input to output a validity score $\hat{v}$, which indicates the likelihood of a sample belonging to the same distribution as the ``real'' samples. The InfoGAN model is structured in a similar manner, however the generator is now a function of both the noise vector $z$ as well as control variables $c$. The control variables may belong to a distribution of choice, but are usually uniformly or categorically distributed. The discriminator model is identical to its GAN counterpart, however it shares some of its layers with an additional auxiliary model $Q(c|x)$. The final layer that is shared between the two models is referred to has a hidden feature layer, which is not directly observed, but serves to contain information about both the validity of the sample as well as its control variable assignments. As such, the auxiliary model outputs a prediction of the control variables $\hat{c}$ just as the discriminator model outputs a prediction of the validity $\hat{v}$. The role of the hidden feature layer is realized through the training process that requires both the validity and control variable predictions to be derived from information contained within the layer.

Additionally, the generator and the discriminator can be conditioned on additional known features. For the Ising configurations, this would be the known physical conditions that the configurations are subject to, namely the temperature and the external magnetic field strength. The inclusion of this information involves conditioning both the generator and discriminator networks on these inputs, denoted with $t$. This type of GAN is referred to as a conditional GAN (CGAN) \cite{cgan}. Through the inclusion of these conditioning variables, training can be stabilized, accelerated, or simply produce better results since the model has access to additional important information regarding the distribution of the training data \cite{cgan}. Furthermore, the conditional distributions are multi-modal, which can mitigate mode collapse in the generator output, though care must be taken to ensure that the generator does not collapse to a single output within the modes provided by the conditional information \cite{cgan_mode}. The cost function reads as

\begin{align}
\min_{G}\max_{D} \mathcal{V}_{\mathrm{CGAN}}(D, G) = &\mathbb{E}_{x\sim P(x)}\qty[\log D(x|t)]+\\\nonumber
&\mathbb{E}_{z\sim P(z)}\qty[\log\qty(1-D(G(z|t)))].
\end{align}

This can then be freely incorporated into the InfoGAN model to produce a so-called InfoCGAN model \cite{infocgan}. The resulting cost is expressed as

\begin{align}
\min_{G}\max_{D} \mathcal{V}_{\mathrm{InfoCGAN}} &= \mathcal{V}_{\mathrm{CGAN}}(D, G) - \lambda I(c; G(c, z|t)) \\
I(c; G(c, z|t)) &= H(c)-H(c|G(c, z|t)) \\\nonumber
&\ge \mathbb{E}_{c\sim P(c), x\sim G(c, z|t)}\qty[\log Q(c|x,t)]+\\\nonumber
&\quad H(c).
\end{align}

\begin{figure}[H]
\centering
\includegraphics[width=0.8\linewidth]{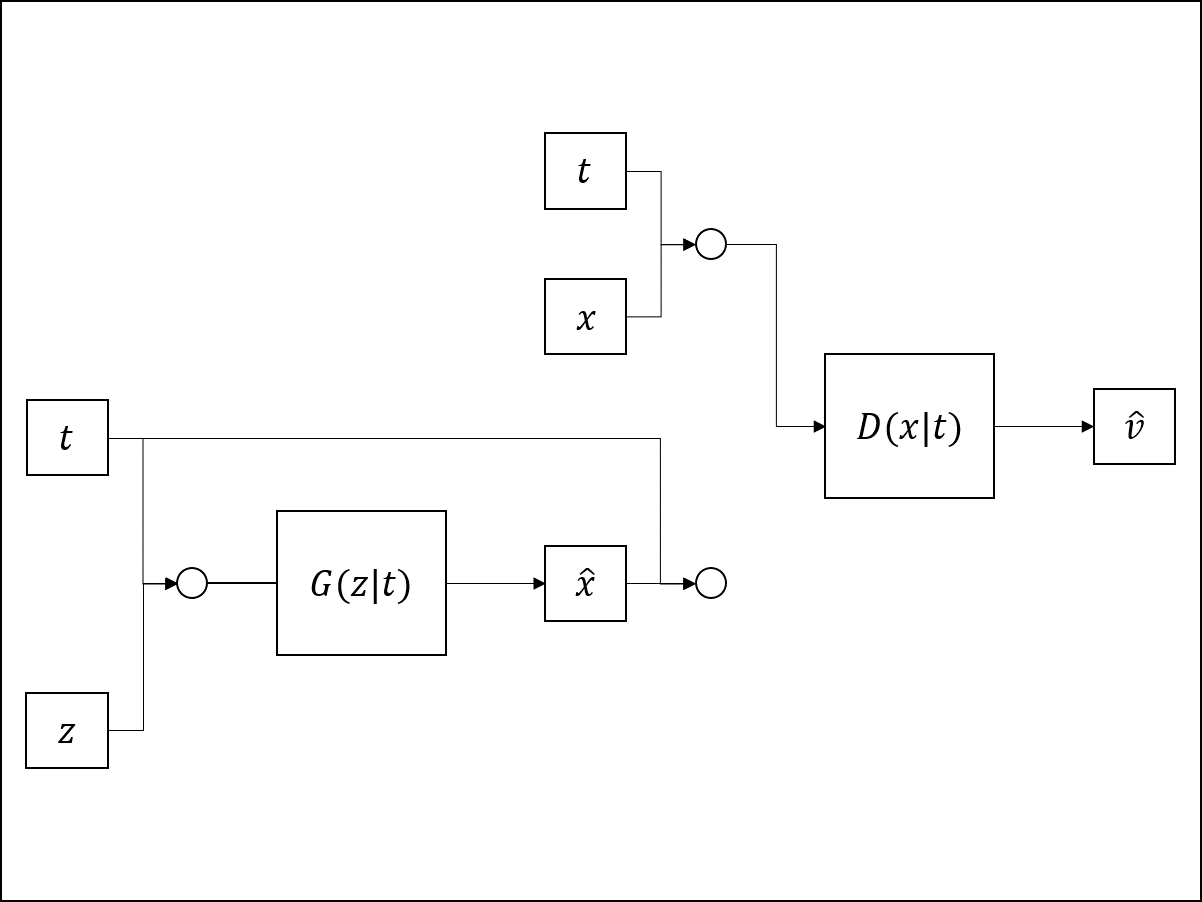}
\caption{\footnotesize This diagram depicts the structure of a CGAN model.}
\label{fig:cgan}
\end{figure}

\begin{figure}[H]
\centering
\includegraphics[width=0.8\linewidth]{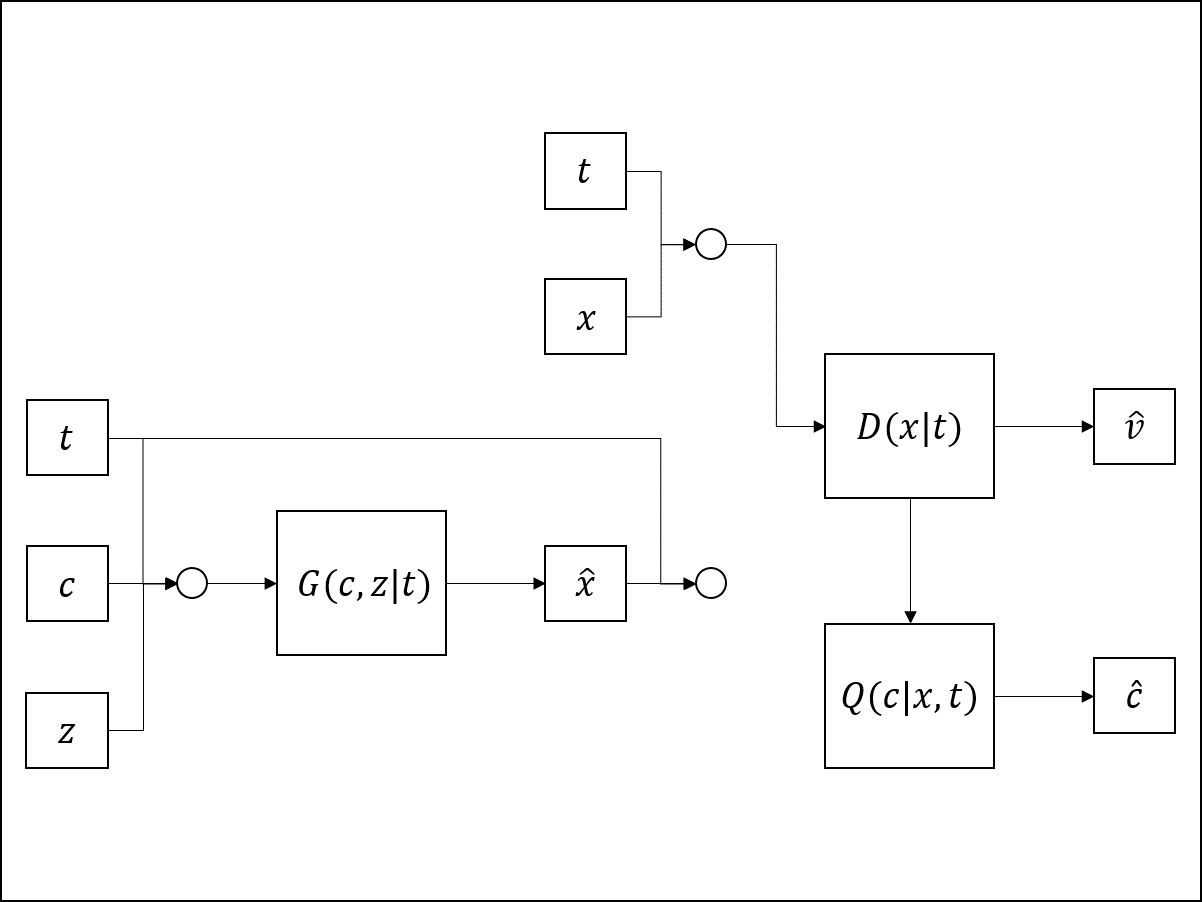}
\caption{\footnotesize This diagram depicts the the structure of an InfoCGAN model.}
\label{fig:infocgan}
\end{figure}

The structures described for the CGAN and InfoCGAN models are respectively depicted in Fig.\ref{fig:cgan} and Fig.\ref{fig:infocgan}. For the CGAN model, the noise vector $z$ as well as the conditional variables $t$ are provided as input to the generator model $G(z|t)$. In addition to outputting a predicted sample $\hat{x}$, the conditional variables $t$ are also fed through as output. The discriminator model $D(x|t)$ additionally takes not just a ``real'' sample $x$ or a ``fake'' sample $\hat{x}$ as input, but also the conditional variables $t$ while outputting the validity prediction $\hat{v}$ as with the normal GAN model. The InfoCGAN model can then be constructed from the CGAN model in a similar way as the InfoGAN model was constructed from the GAN model through the inclusion of the control code $c$. The generator and auxiliary networks respectively become $G(c,z|t)$ and $Q(c|x,t)$.

\section{Methods}

Monte Carlo simulations of the 2-dimensional square Ising model with linear size $l=27$ were performed over a range of both temperatures and external fields through spin-flip attempts subject to a Metropolis criterion. A large number of configurations were collected for each combination of temperature and external field following thermal equilibration. After the configurations are collected, the Ising spins are rescaled to such that $\Bqty{-1, +1} \rightarrow \Bqty{0, 1}$, which is a common approach in data science when considering binary-valued features. Physically, this is consistent with a lattice gas model, which is equivalent to the Ising model.

These configurations obtained through Monte Carlo sampling compose the training data for an InfoCGAN model that learns to both classify these Ising configurations and generate convincing Ising configurations given a random vector input, a temperature and external field, and a classification. The trained auxiliary model is then used to classify the Ising configurations from the training set and the classifications are analyzed as functions of the temperature and external field to demonstrate a correspondence with the expected physical phases of the Ising model across the temperature and external field range. Five categorical control variables were used for the classifications, which are intended to correspond with the expected three distinct physical phases (ferromagnetic spin-up, ferromagnetic spin-down, and paramagnetic) as well as two classifications that consist of configurations that show still ferromagnetic ordering, albeit with some spins flipped. These two classifications are expected to contain ferromagnetic configurations that are not perfectly (or near-perfectly) ordered due to the finite probability of spins flipping in the Monte Carlo simulations in addition to configurations associated with crossover phenomena in the presence of an external field. These additional classifications were found to be necessary for reliably reproducing the categorical assignments, as using only three categorical assignments would would result in these configurations either being absorbed into the classifications corresponding to perfectly (or near perfectly) ordered ferromagnetic configurations or the classification corresponding to the high temperature weak field configurations. Sometimes, the training process would even predict these categorical assignments in a non-symmetric manner across the vanishing field line if only three categories were used.

During training, the generator and the discriminator reach a Nash equilibrium in which neither the discriminator or the generator gain advantages through continuing their adversarial ``game.'' Once the classifications of the Ising configurations are obtained from the auxiliary network, the average classifications with respect to temperature and the applied external field are compared with the expected behavior of the Ising model with respect to these physical conditions. Along the vanishing field line $(H = 0)$, the average classification of the Ising configurations into the class corresponding to high temperature samples as a function of temperature is interpreted as a probability of the Ising configurations belonging to the paramagnetic phase as a function of temperature. From this probability, an estimation of the critical point is obtained and compared to the behavior of the magnetic susceptibility and the specific heat capacity as functions of temperature. Correspondence of the estimated critical point derived from the auxiliary classifications to the peaks in these functions indicates correspondence between the paramagnetic phase and the class corresponding to the high temperature configurations obtained from the auxiliary network.

\section{Results}

The loss history of the InfoCGAN does indicate the tendency towards a Nash equilibrium, though with occasional classification errors in the later epochs that are nonetheless statistically conscionable. The discrimination loss is essentially equivalent for the ``real'' and ``fake'' samples and is consistently lower than the generation loss, but the results are well within expectations. Training stability was rather quickly obtained. All of the plots in this section are generated with the MatPlotLib package using a perceptually uniform colormap \cite{matplotlib}.

\begin{figure}[H]
\centering
\includegraphics[width=0.8\linewidth]{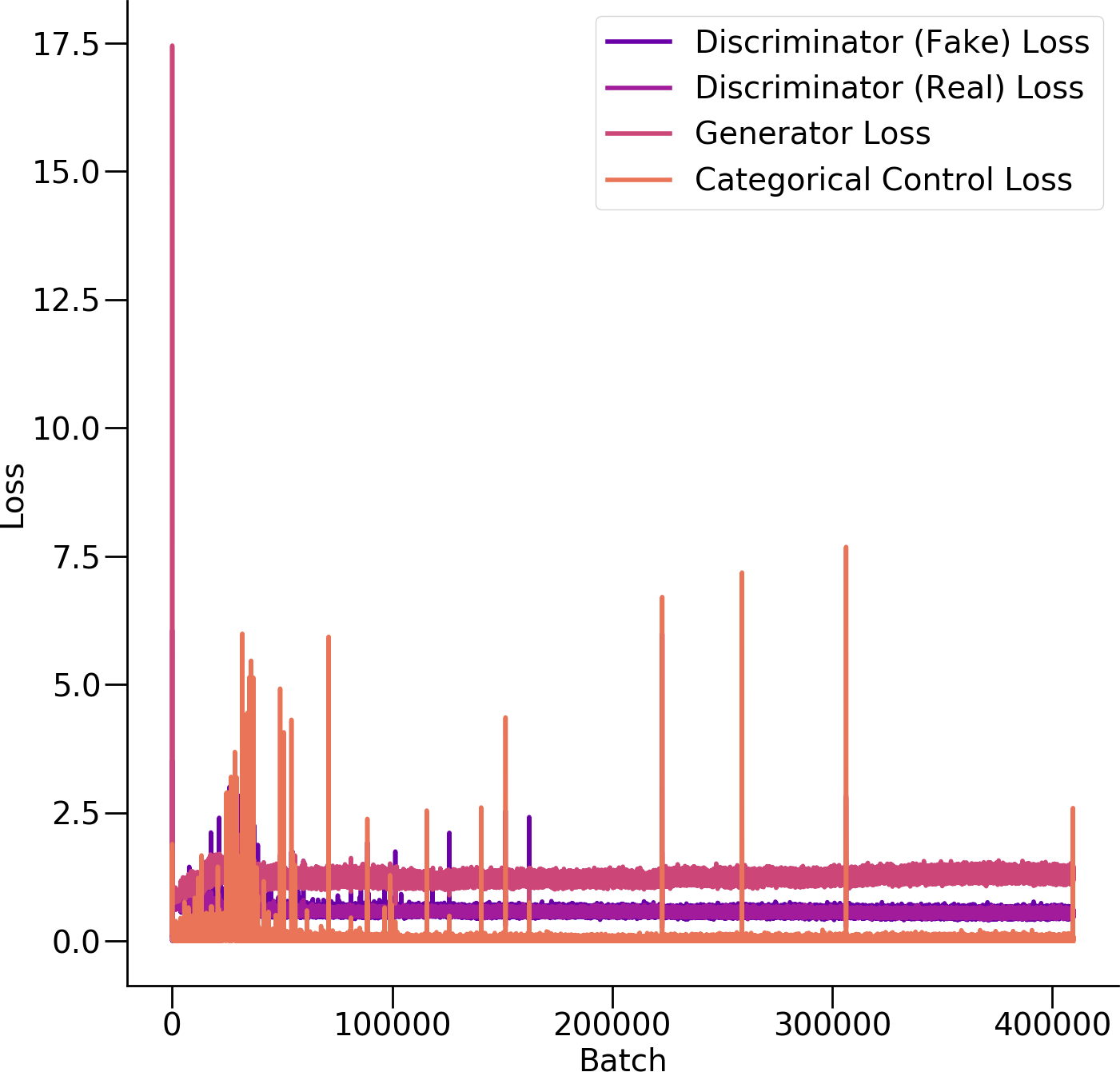}
\caption{\footnotesize This diagram depicts the average batch loss history of the InfoCGAN model trained on the Ising configurations. The discriminator and generator losses are the validity predictions made by the discriminator while respectively training the discriminator and the generator. The categorical control loss is the categorical crossentropy loss of the categorical control code predictions made by the auxiliary network while training the generator.}
\label{fig:loss_batch}
\end{figure}

\begin{figure}[H]
\centering
\includegraphics[width=0.8\linewidth]{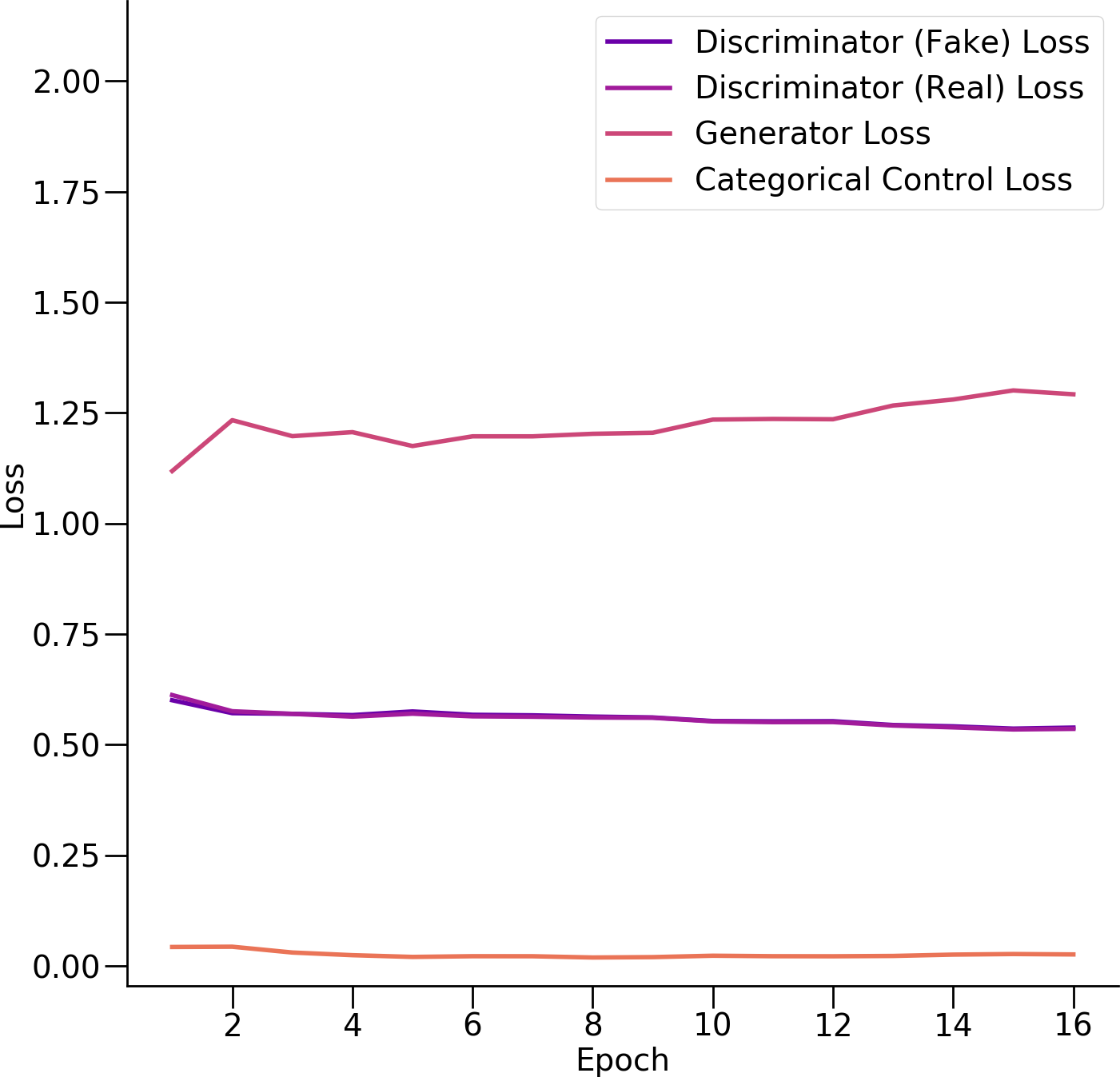}
\caption{\footnotesize This diagram depicts the epoch loss history of the InfoCGAN model trained on the Ising configurations. The discriminator and generator losses are the validity predictions made by the discriminator while respectively training the discriminator and the generator. The categorical control loss is the categorical crossentropy loss of the categorical control code predictions made by the auxiliary network while training the generator.}
\label{fig:loss_epoch}
\end{figure}

The loss history with respect to the batches is depicted in Fig.\ref{fig:loss_batch} while the loss history with respect to the epochs is depicted in Fig.\ref{fig:loss_epoch}. The batch losses refer to the losses for each individual minibatch used during training, which consist of subsets of the total training set. The epoch losses are then the average losses across all samples in the training set for each epoch, in which every sample in the training set has had an opportunity to update the model parameters. The discriminator losses, both ``real'' and ``fake,'' are binary crossentropies of between the validity predictions made by the discriminator model and the provided target validities while it is respectively trained on samples from the training set and samples from the generator model. The loss for the generator is also the binary crossentropy between the validity prediction provided by the static discriminator model and the provided target validities while the generator and auxiliary models are trained. The categorical control loss is then the categorical crossentropy of the predicted categorical controls from the auxiliary model with respect to the categorical codes provided to the generator model during training of the generator and auxiliary models.

\begin{figure}[H]
\centering
\includegraphics[width=0.8\linewidth]{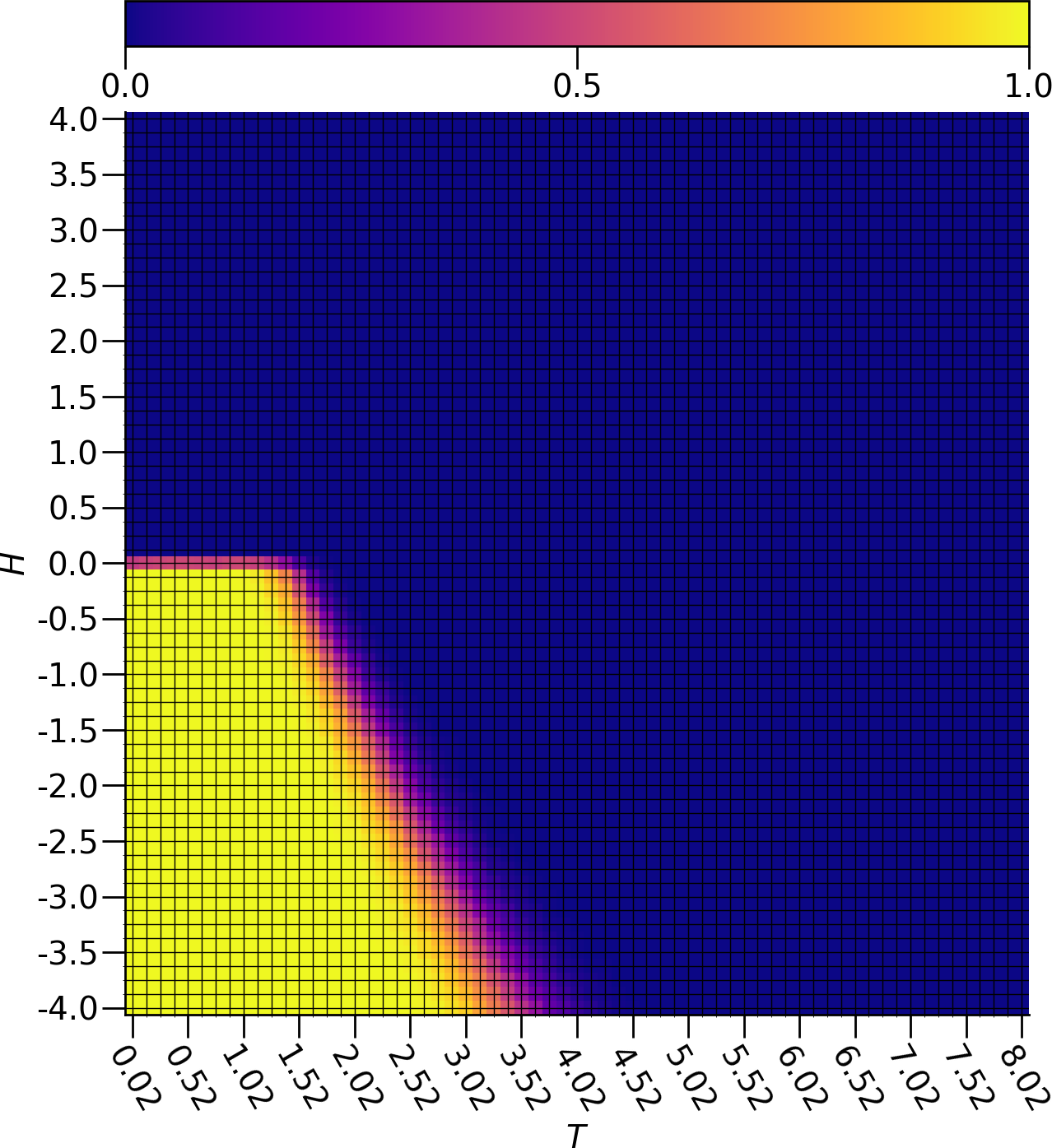}
\caption{\footnotesize This diagram depicts the average classification probability for the first categorical variable with respect to temperature and the external magnetic field.}
\label{fig:infocgan_0}
\end{figure}

\begin{figure}[H]
\centering
\includegraphics[width=0.8\linewidth]{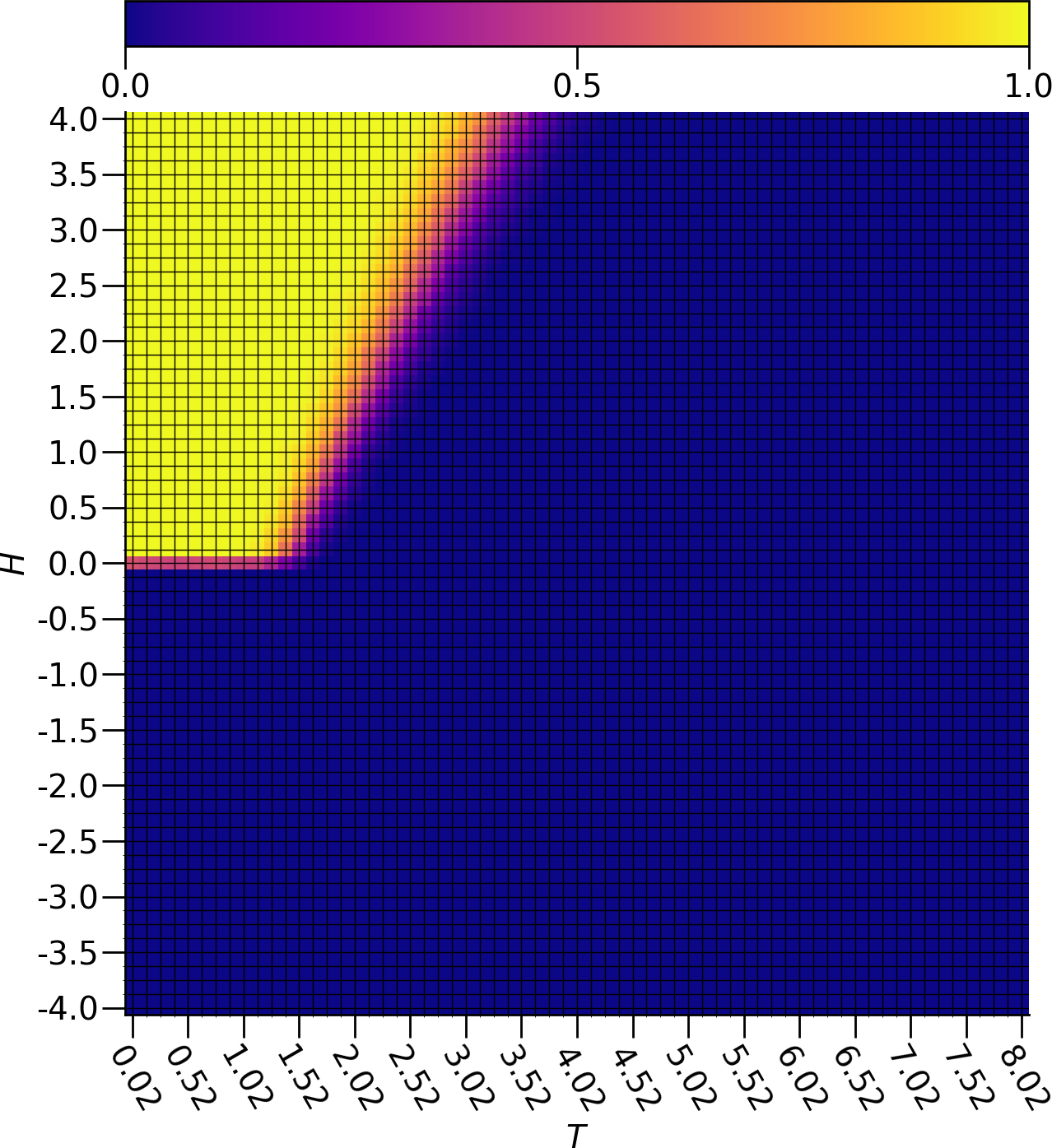}
\caption{\footnotesize This diagram depicts the average classification probability for the second categorical variable with respect to temperature and the external magnetic field.}
\label{fig:infocgan_1}
\end{figure}

The predicted classification probabilities for the first two categorical control variables are shown with respect to the temperature and the external magnetic field for the Ising configurations are depicted in Fig.\ref{fig:infocgan_0} and Fig.\ref{fig:infocgan_1}. The diagrams illustrate the probability $P_{S_i}(H, T)$ that samples located at an $(H, T)$ coordinate belongs to class $S_i$, where Fig.\ref{fig:infocgan_0} and Fig.\ref{fig:infocgan_1} are for classes $S_0$ and $S_1$ respectively. The color bar defines what is meant by the coloration, where the darkest regions exhibit the smallest probability $P_{S_i}(H, T) = 0$ and the brightest regions exhibit the largest probability $P_{S_i}(H, T) = 1$. The results are exceptional, as the regions of high classification probabilities reflect the expected locations of the low-temperature ordered ferromagnetic phases, complete with symmetry across the vanishing field line. The classification probabilities are very close to 0.5 for each category along the low-temperature regime of the vanishing field, reflecting the fact that in the absence of an external field, there is no imposed preference on the type of magnetic order exhibited by the system. The boundaries of these high probability regions for these categories are rather sharply defined. This is encouraging, as both the training Ising configurations and the generated configurations for these regions possess consistent thermodynamics, showing an average magnetization extremely close to either -1 or +1 across the regions depending on the external field (excluding the vanishing field case, of course). Outside of these predicted classification regions, there is a consistent departure from the perfectly (or near-perfectly) ordered states caused by thermal fluctuations.

\begin{figure}[H]
\centering
\includegraphics[width=0.8\linewidth]{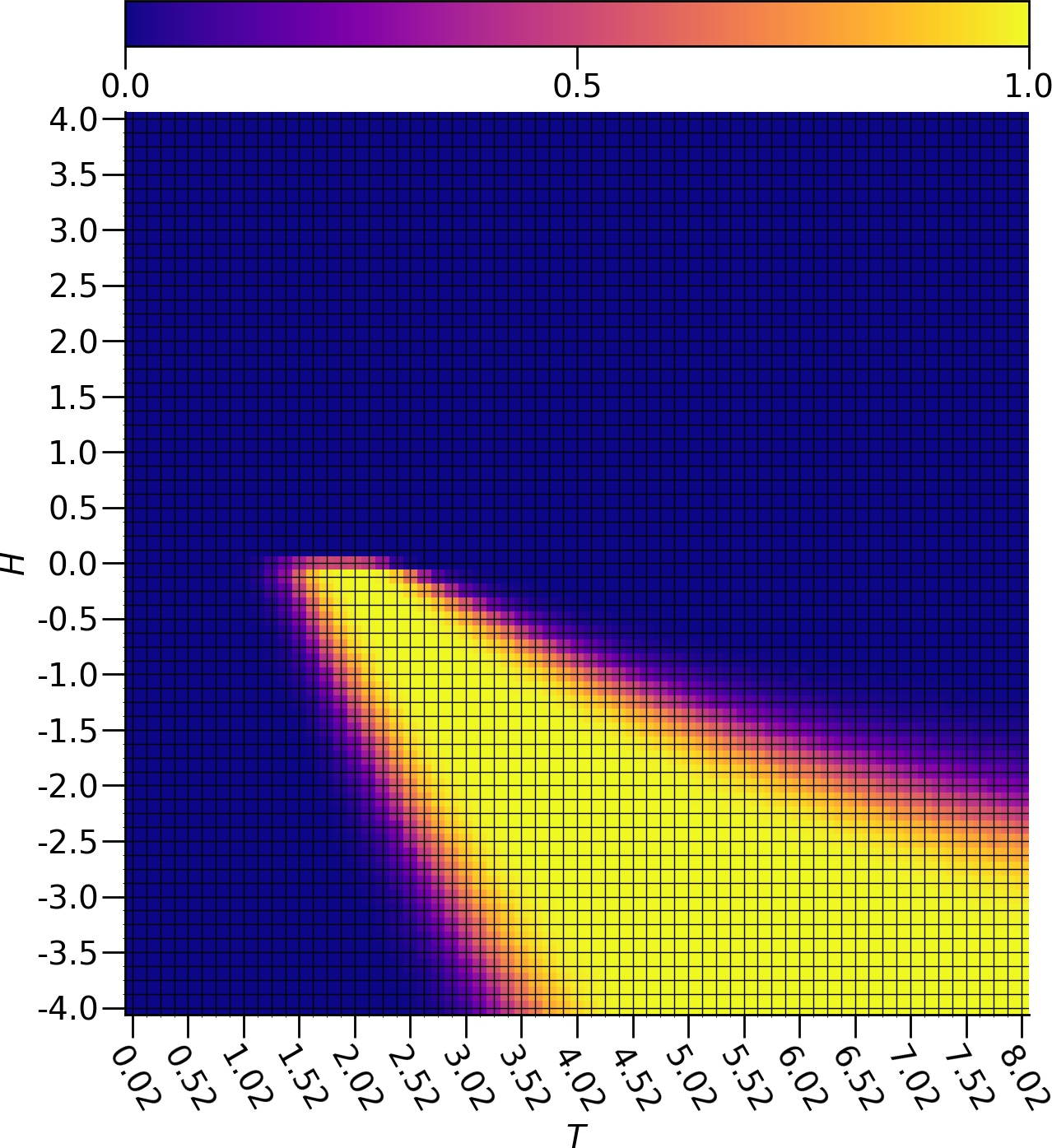}
\caption{\footnotesize This diagram depicts the average classification probability for the third categorical variable with respect to temperature and the external magnetic field.}
\label{fig:infocgan_2}
\end{figure}

\begin{figure}[H]
\centering
\includegraphics[width=0.8\linewidth]{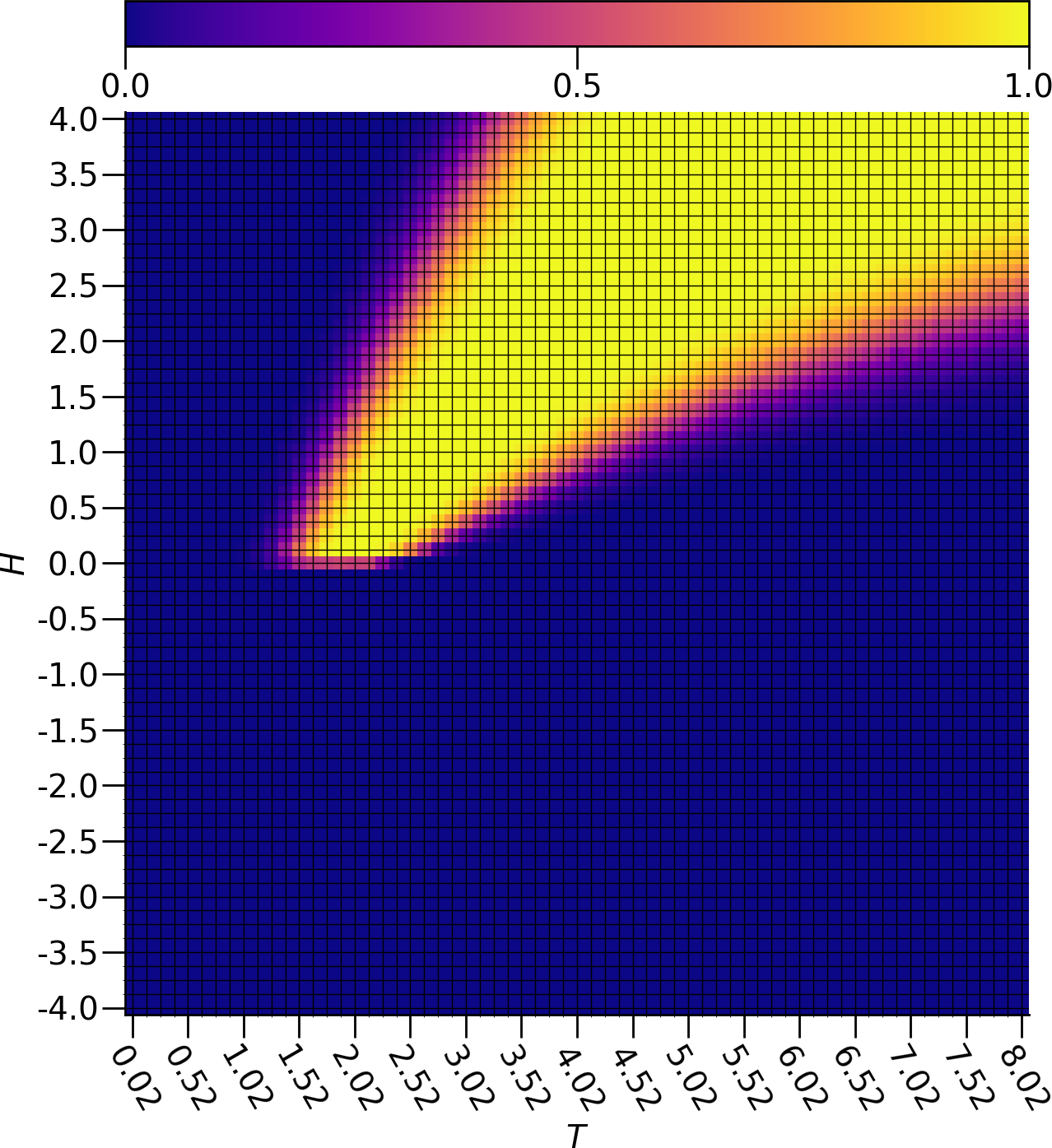}
\caption{\footnotesize This diagram depicts the average classification probability for the fourth categorical variable with respect to temperature and the external magnetic field.}
\label{fig:infocgan_3}
\end{figure}

The predicted classification probabilities for the next two categorical control variables are shown with respect to the temperature and the external magnetic field for the Ising configurations are depicted in Fig.\ref{fig:infocgan_2} and Fig.\ref{fig:infocgan_3}. These classification regions capture the aforementioned configurations in which there are significant departures from the completely ordered ferromagnetic states, but the nearest-neighbor interactions and interactions with the external field (where applicable) are still strong enough that ferromagnetic ordering is readily apparent. Investigation into the configurations contained in these regions shows a smooth decay of the average magnetization with respect to increasing temperature in the presence of an external field, consistent with the expected characteristics of crossover phenomena. Furthermore, as the external field strength increases, the region shifts to greater temperatures, consistent with the expectation that the higher temperatures are required for thermal fluctuations to overpower the influence imposed by the external field. Once again, there is an observed symmetry of the two regions across the vanishing field line as well as reasonably sharply defined region boundaries, similar to what was observed with the near perfectly ordered ferromagnetic regions. However, note that there are non-zero probabilities for these classifications along the vanishing field line. In the thermodynamic limit as system size tends towards infinite $(N \rightarrow \infty)$, this is not a valid result. This behavior can be described as an artifact of the machine learning approach in which the configurations that are approaching the critical temperature demonstrate diminishing ferromagnetic ordering that is qualitatively similar to the properties of the configurations associated with crossover in the presence of an external field. It is expected that this behavior would diminish with increasing system size, but that is beyond the scope of this study at present time.

The crossover line can be roughly be defined as the line in a parameter space for which the correlation length driven by two relevant parameters, in this case the external field and temperature, are roughly comparable with one other. One pragmatic definition is the Widom line, which can be defined as the locus for the maximum of the heat capacity.

The obtained finite area, as opposed to a sharp line, is a reflection of the fact that the present machine learning approach does not pinpoint a sharp single line to separate the two regimes. Instead, it characterizes the line through the competition between the two relevant parameters. The growth of the area to encompass a larger temperature range as the external field grows in magnitude is perhaps a reflection of the weaker correlation at high temperature which leads to a larger uncertainty. This is perhaps unsurprising, as drawing the end of a crossover line is difficult and often not well defined.

\begin{figure}[H]
\centering
\includegraphics[width=0.8\linewidth]{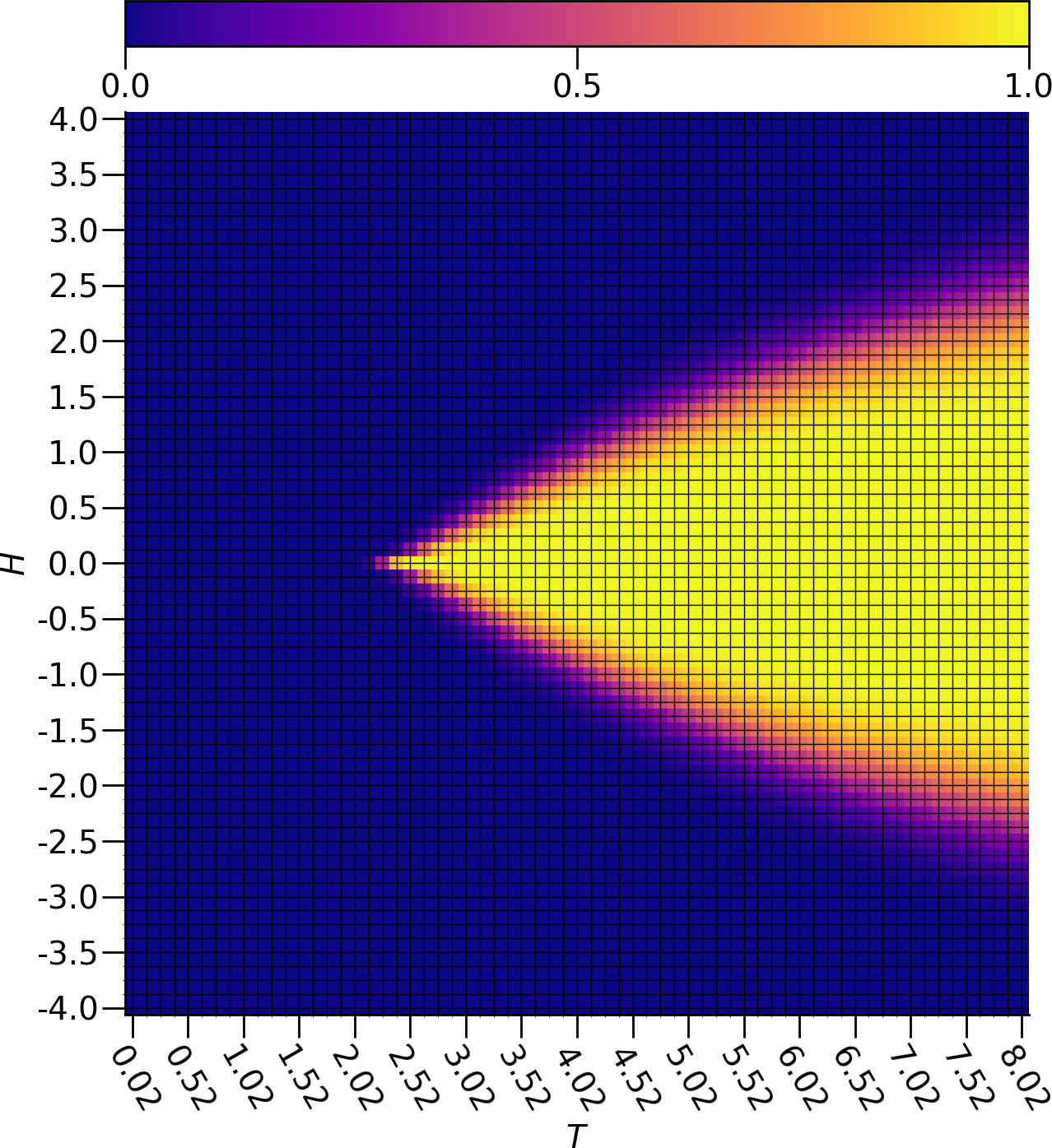}
\caption{\footnotesize This diagram depicts the classification probability for the fifth categorical variable with respect to temperature and the external magnetic field.}
\label{fig:infocgan_4}
\end{figure}

The predicted classification probability for the final categorical control variable with respect to the temperature and the external magnetic field for the Ising configurations is depicted in Fig.\ref{fig:infocgan_4}. This classification region is corresponds well to configurations at high temperature under the influence of a weak external field, as the configurations contained by it exhibit nearly zero average magnetization due to the effects of thermal fluctuations. Where the ferromagnetic regions failed to predict the critical temperature due to the artifacts of the machine learning categorization approach, this region compensates, as it begins in the immediate vicinity of the true critical point of $T_C \approx 2.269$ calculated by Kramers-Wanier duality \cite{kwd,kwd_2,kwd_3}. Consistent with the prior two sets of classification regions, there is a symmetry across the vanishing field. By contrast with the others, however, there is no need to partition the classification across the vanishing field as the thermal fluctuations are overpowering the influence of the external field and thus the configurations are only weakly responding to it.

In order to further investigate how well-approximated the critical point actually is, a logistic curve can be fit using orthogonal distance regression to the average classification probabilities of the category corresponding to the paramagnetic phase with respect to temperature along $H = 0$. This take the form $P_{S_{\mathrm{paramagnetic}}}(T) = \frac{1}{1+e^{-k(T-\hat{T}_C)}}$ where $k$ is the logistic growth rate of the probability and $\hat{T}_C$ is the estimated critical temperature.

\begin{figure}[H]
\centering
\includegraphics[width=0.8\linewidth]{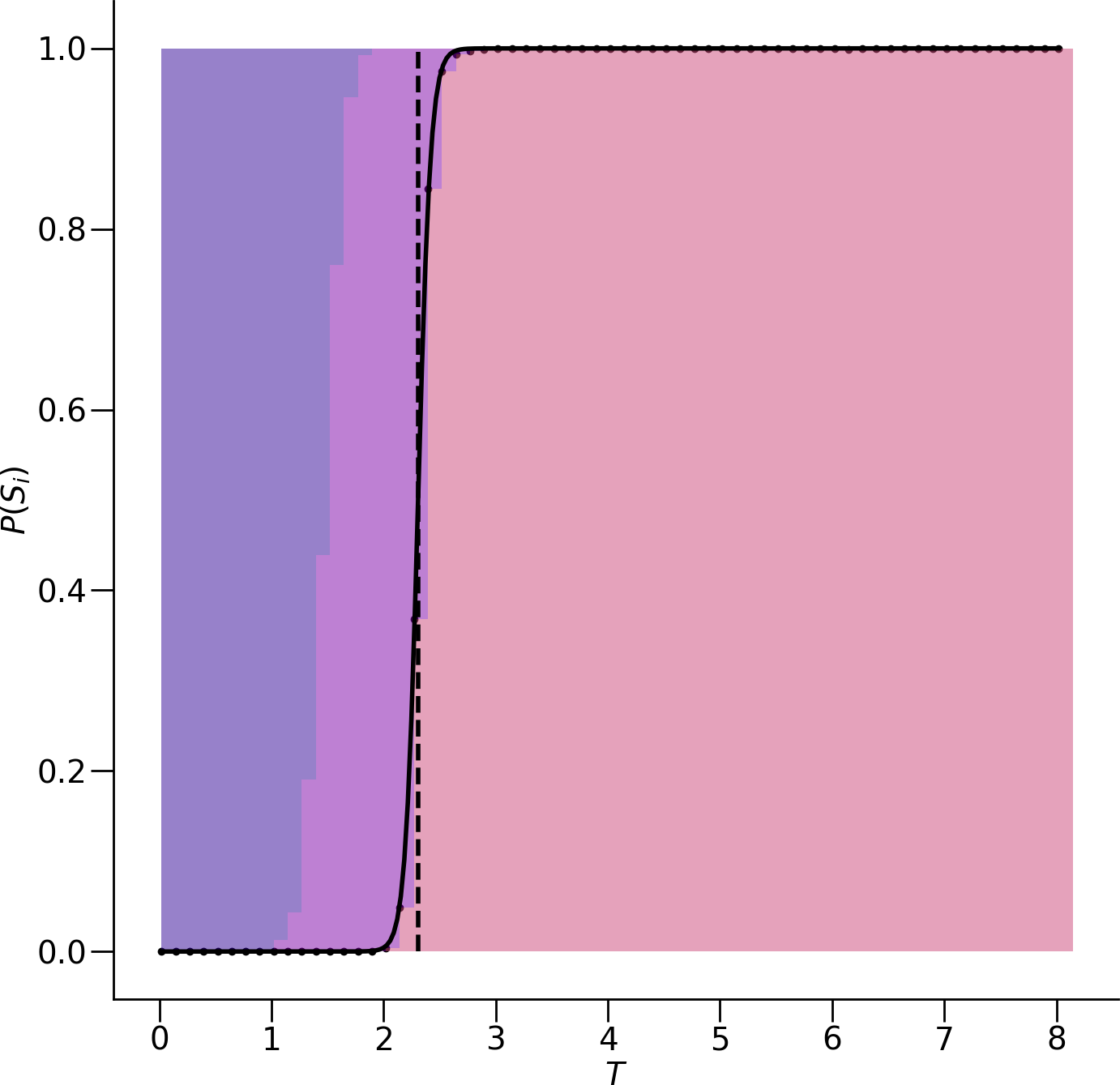}
\caption{\footnotesize This diagram depicts the average classification probabilities of the class assignments predicted by the auxiliary network with respect to temperature in the vanishing field case. Deep purple represents the classifications corresponding to configurations with near perfect ferromagnetic ordering, light purple represents the classifications corresponding to configurations with diminished ferromagnetic ordering, and pink represents the classifications corresponding to disordered configurations. The logistic curve is fit to the disordered probabilities and provides a critical temperature estimate of $\hat{T}_C \approx 2.308$.}
\label{fig:state}
\end{figure}

This is shown in Fig.\ref{fig:state}, where the dotted line indicates the critical temperature estimate at $\hat{T}_C \approx 2.308$. This provides an overestimate of $\approx 1.688\%$. Due to the finite-size effects caused by the rather small system size of linear size $l = 27$, an overestimate of the critical temperature is expected, so this result is consistent with expectations.

\begin{figure}[H]
\centering
\includegraphics[width=0.8\linewidth]{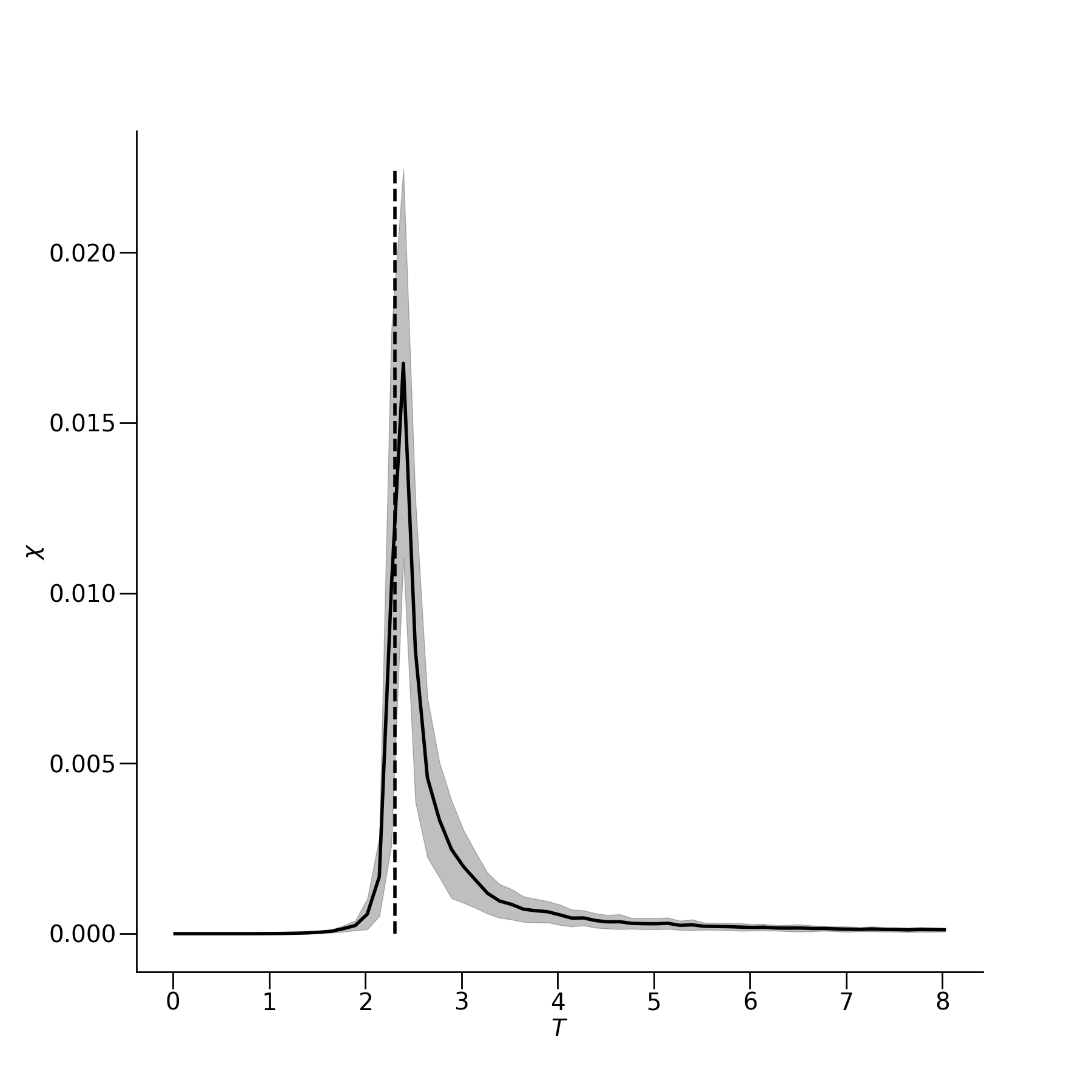}
\caption{\footnotesize This diagram depicts the magnetic susceptibility of the Ising configurations with respect to temperature in the vanishing field case. The dotted line corresponds to the estimated critical temperature of $\hat{T}_C \approx 2.308$. The grey regions correspond to the Monte Carlo error.}
\label{fig:ms}
\end{figure}

\begin{figure}[H]
\centering
\includegraphics[width=0.8\linewidth]{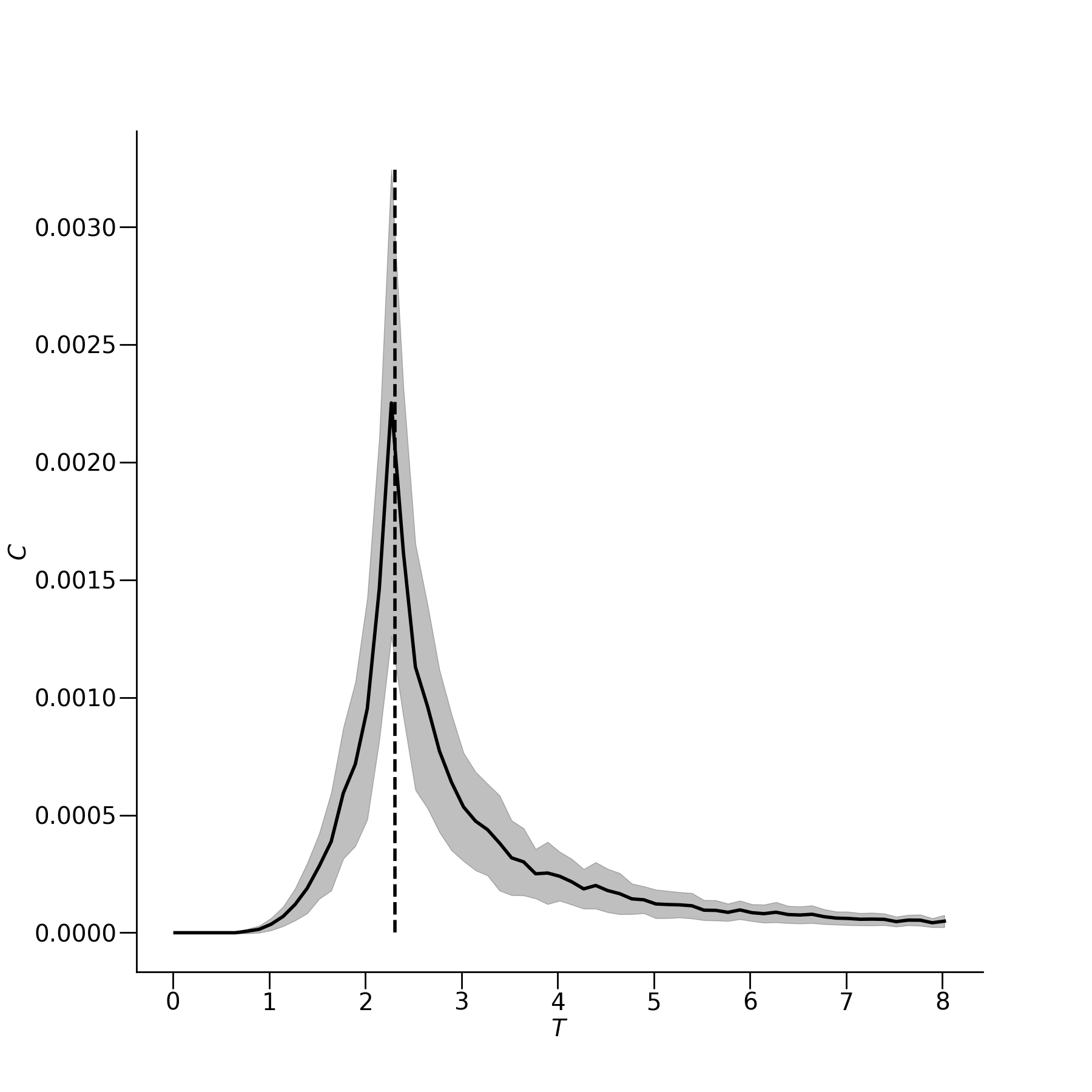}
\caption{\footnotesize This diagram depicts the specific heat capacity of the Ising configurations with respect to temperature in the vanishing field case. The dotted line corresponds to the estimated critical temperature of $\hat{T}_C \approx 2.308$. The grey regions correspond to the Monte Carlo error.}
\label{fig:sh}
\end{figure}

This estimation of the critical point is consistent with the peaks observed in the magnetic susceptibility $\chi$ and the specific heat capacity $C$ respectively depicted in Fig.\ref{fig:ms} and Fig.\ref{fig:sh}, which would be divergent in the thermodynamic limit. The Monte Carlo errors are calculated using the jackknife technique and are depicted with the grey regions in the figures. Note that the errors are more pronounced in the vicinity of the critical temperature, which accounts for the inconsistency of the location of the peaks between the two measured quantities. These errors are expected to be much lower with larger system sizes. This provides significant evidence of the consistency of the critical temperature estimation provided through the auxiliary network classification with the known physics.

\begin{figure}[H]
\centering
\includegraphics[width=0.8\linewidth]{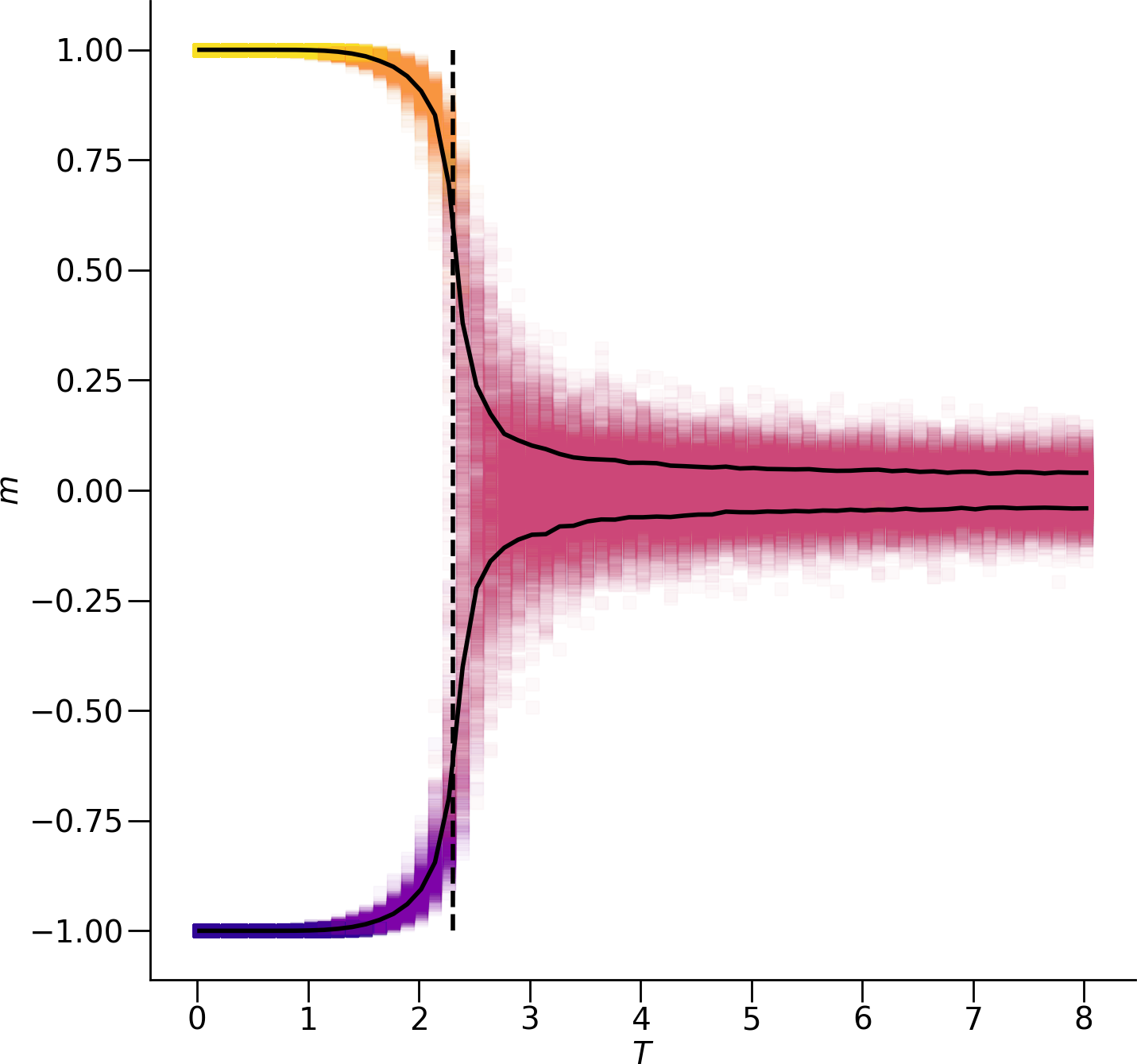}
\caption{\footnotesize This diagram depicts the average magnetizations of the Ising configurations with respect to temperature in the vanishing field case, separated into plurality spin-up and spin-down varieties. The samples are colored according to their classification, with yellow and blue representing the classifications corresponding to configurations with near perfect ferromagnetic ordering, orange and purple representing the classifications corresponding to configurations with diminished ferromagnetic ordering, and pink representing the classifications corresponding to disordered configurations. The dotted line corresponds to the estimated critical temperature of $\hat{T}_C \approx 2.308$.}
\label{fig:magnetization}
\end{figure}

\begin{figure}[H]
\centering
\includegraphics[width=0.8\linewidth]{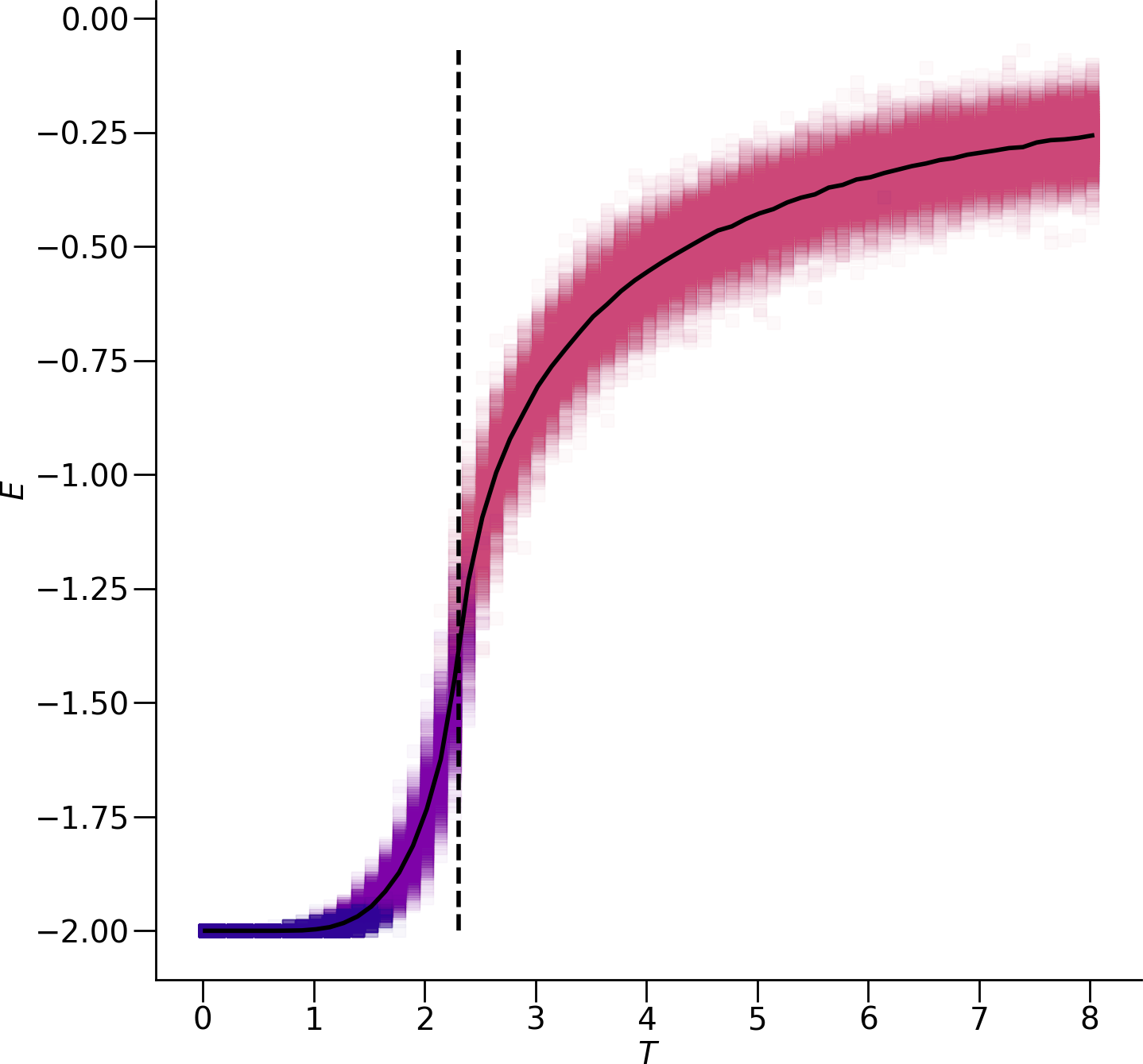}
\caption{\footnotesize This diagram depicts the average energies of the Ising configurations with respect to temperature in the vanishing field case. The spin-up and spin-down varieties of the classifications have been combined. The samples are colored according to their classification, blue representing the classifications corresponding to configurations with near perfect ferromagnetic ordering, purple representing the classifications corresponding to configurations with diminished ferromagnetic ordering, and pink representing the classifications corresponding to disordered configurations. The dotted line corresponds to the estimated critical temperature of $\hat{T}_C \approx 2.308$.}
\label{fig:energy}
\end{figure}

The magnetizations $m$ and energies $E$ of the Ising configurations with respect to temperature are additionally respectively shown in Fig.\ref{fig:magnetization} and Fig.\ref{fig:energy}. The predicted critical temperature is consistent with expectations, as it is located where the strict separation of the spin-up and spin-down configurations breaks down for the magnetization and where the inflection occurs in the energy.

We emphasize that the present analysis is based on one system size. Performing proper finite-size scaling necessarily involves a proper characterization of the error from the machine learning approach, which is not well understood. For the conventional finite-size scaling approach, the error can be argued through various fitting criteria for the finite-size scaling ansatz. There is no simple analogy to this in the machine learning approach. 

\section{Conclusions}

The InfoCGAN auxiliary classification network has performed extremely well in predicting phases of the 2-dimensional Ising model. It is important to note how well the approach agrees with known properties of the Ising model despite having no access to any physical \textit{a priori} information beyond the raw configurations, the temperatures, and the external field strengths. From this approach, classifications are obtained that correspond to configurations with near perfect ferromagnetic ordering, diminished ferromagnetic ordering, and disorder. Furthermore, an exceptional estimation of the critical point is provided by this approach given the influence of finite-size effects.

These results show encouraging results to motivate further development of approaches to machine learn phases of matter. For many physical systems, order parameters have not been identified through conventional means and the phase diagrams elude discovery. However, simulation data consisting of configurational information with respect to thermal parameters is readily available. It is clear from this work that machine learning approaches are indeed capable of determining a structural criterion to partition physical samples into classes that correspond to different phases of matter and are even flexible enough to allow for conditioning of the model on additional physical parameters that may prove useful to that end. The use of the InfoCGAN model to perform this task provides a particular advantage in providing a direct unsupervised classification scheme of Ising configurations into classes that correspond to physical phases whereas other research into unsupervised machine learning of the Ising model has instead mapped configurations to representations that can be later used as classification criteria. The direct unsupervised classification approach is applicable to situations in which latent representations may not be easily translated to classification criteria.

There are many opportunities beyond investigating more complex systems by introducing improvements to this method beyond the scope of this work. For instance, finite-size scaling is an important approach towards addressing limitations presented by finite-sized systems for investigation critical phenomena \cite{fsc}. Establishing correspondence between the categorical codes of different system sizes is a challenging proposition, as different InfoCGAN models will need to be trained for each system size, which in turn may require different hyperparameters and training iteration counts to provide similar results. Consequently, numerical difficulties can arise when performing finite-size scaling analysis, as the variation of predicted properties with respect to system size may be difficult to isolate from the variation of systemic errors due to different neural networks being used to extract said properties. Nevertheless, this would be a significant step towards improving InfoCGAN classification of physical configurations into classes that correspond to physical phases.

\section{Acknowledgments}

This work is funded by the NSF EPSCoR CIMM project under award OIA-1541079. Additional support (KMT) was provided by NSF Materials Theory grant DMR-1728457. An award of computer time was provided by the INCITE program. This research also used resources of the Oak Ridge Leadership Computing Facility, which is a DOE Office of Science User Facility supported under Contract DE-AC05-00OR22725.

\appendix

\section{Simulation and Model Parameters}

The simulated Ising configurations obtained are generated using a typical Monte Carlo simulation across a large range of temperatures and external magnetic fields using a Metropolis criterion for spin-flip attempts. This is implemented using the Python programming language with NumPy operations, Numba JIT compilation, and Dask parallelization \cite{python,numpy,numba,dask}. The lattice size chosen for the square 2-dimensional configurations was $l = 27$. A single spin flip attempt consists of flipping the spin of a single lattice site, calculating the resulting change in energy $\Delta E$, and then using that change in energy to define the Metropolis criterion $\exp(-\frac{\Delta E}{T})$. If a randomly generated number is smaller than said Metropolis criterion, the configuration resulting from the spin-flip is accepted as the new configuration. The data analyzed in this work consists of 1,024 square Ising configurations of side length 27 with periodic boundary conditions across 65 external field strengths and 65 temperatures respectively sampled from $\qty[-4, 4]$ and $\qty[0.02, 8.02]$ on uniformly spaced grids. The interaction strength was set to unity such that $J_{ij} = J = 1$. Each sample was equilibrated with 8,192 Monte Carlo updates before data collection began. Data was then collected at an interval of 8 Monte Carlo updates for each sample up to a sample count of 1,024. At the end of each data collection step, a replica exchange Markov chain Monte Carlo move, also known as parallel tempering, was performed across the full temperature range for each set of Ising configurations that shared the same external field strength \cite{remcmc,par_temp_1,par_temp_2}.

In this work, the  Ising spins are rescaled such that a spin-down atomic spins carry the value 0 and spin-up atomic spins carry the value 1, which is a standard setup for binary-valued features in data science applications. This representation is consistent with the lattice gas model, which is equivalent to the Ising model.

The InfoCGAN model is trained on the collection of Ising configurations using five categorical control variables and two conditional variables, taking the values of the temperature and the applied external magnetic field. The temperatures are normalized to a range of $\qty[0, 1]$ and the external magnetic fields are normalized to a range of $\qty[-1, +1]$. The categorical control variables, conditional variables, and a Gaussian latent space of size 128 are concatenated to produce the latent input vector for the generator network. Every hidden layer uses LeCun normal kernel initializations with scaled exponential linear unit (SELU) activations such that the networks are self-normalizing networks (SNNs) \cite{snn}. Due to the 2-dimensional structure of the Ising configurations, it is a natural choice to implement the InfoCGAN as a convolutional neural network (CNN) \cite{cnn}. The neural network itself was implemented in the Python language using the TensorFlow framework with Keras layers \cite{python,tf,keras}.

The first hidden layer in the generator network is a dense layer of size 256, which is then reshaped to a size of (1, 1, 256). Two transposed convolutions are then iteratively applied with a number of filters that decreases by a factor of 2 upon each iteration. A final transposed convolution is applied with a single filter such that an output tensor of shape (27, 27, 1) is designated as an output layer subject to a logistic activation function with output restricted to the interval $[0, 1)$, which can be treated as the probability of a spin-up lattice site. All of the transposed convolution layers are implemented with a kernel shape of (3, 3) and a stride of (3, 3). The input conditional variables are additionally fed forward as output.

The discriminator network takes an Ising configuration of shape (27, 27, 1) as well as conditional variables as input. The first component of the discriminator network performs iterative convolutions on the input Ising configuration, which is the inverse of the iterative transposed convolutions performed by the generator network. The final convolution produces an output shape of (1, 1, 256) which is then flattened to remove the vestigial dimensions. This is treated as a hidden latent space containing structural information about the input Ising configurations extracted by the convolutional layers. Up until this point, all of the layers in the discriminator network are shared with the auxiliary network such that all of the weights and biases are shared. After this point, the discriminator and auxiliary networks will indeed possess similar structure, but they will be allowed to optimize the weights and biases of these final layers such that they are allowed to respond to different features in the hidden latent space to optimize their respective predictions. The discriminator and auxiliary networks then independently feed the hidden latent space output into dense layers of size 135. The resulting output is concatenated with the input conditional variables and respectively fed into the output layers for each network. The output layer of the discriminator network is a dense layer of size 1 implements with a logistic activation function. By contrast, the output layer of the auxiliary network is a dense layer of size 5 implemented with a softmax activation function.

The loss function is implemented with binary crossentropy loss for the validity score and categorical crossentropy for the categorical control predictions. Stochastic gradient descent (SGD) is used as the optimizer for the discriminator with a learning rate of 0.01 and adaptive moment estimation (Adam) is used as the optimizer for the generator with $\beta_1 = 0.5$ with a learning rate of 0.001 in order to prevent the discriminator from overpowering the generator during learning \cite{sgd,adam}. A value of $\lambda = 1$ is employed for the mutual information loss from the auxiliary network.

Training is performed for sixteen epochs with a batch size of 169, resulting in 25,600 batches for each epoch. After sufficient training, the auxiliary network can be used to predict the classification probabilities of the Ising configurations with respect to the five categories. If one would desire it, the generator network can additionally be used to generate new convincing Ising configurations given a categorical assignment, temperature, external magnetic field, and a random Gaussian latent vector.

\bibliography{refs}

\end{document}